\documentclass[twocolumn,dvipsnames,twocolappendix]{aastex631}
\usepackage{amsmath,amsfonts,amssymb,graphicx,multirow}
\usepackage{mathrsfs}
\usepackage[]{hyperref}
\hypersetup{colorlinks=true}

\usepackage{booktabs}
\usepackage{amsmath}
\usepackage{empheq}

\begin{document}
\title{Relativistic outflows power a quasi-periodic eruption: constraints on energetics, mass loss, and emission mechanisms}

\author[0000-0002-0568-6000]{Joheen Chakraborty}
\thanks{joheen@mit.edu}
\affiliation{Department of Physics \& Kavli Institute for Astrophysics and Space Research, Massachusetts Institute of Technology, Cambridge, MA 02139, USA}

\author[0000-0003-0172-0854]{Erin Kara}
\affiliation{Department of Physics \& Kavli Institute for Astrophysics and Space Research, Massachusetts Institute of Technology, Cambridge, MA 02139, USA}

\author[0000-0002-1568-7461]{Wenbin Lu}
\affiliation{Department of Astronomy \& Theoretical Astrophysics Center, University of California, Berkeley, CA 94720, USA}

\author[0000-0002-4670-7509]{Brian D. Metzger}
\affiliation{Department of Physics and Columbia Astrophysics Laboratory, Columbia University, New York, NY 10027, USA}
\affiliation{Center for Computational Astrophysics, Flatiron Institute, 162 5th Ave, New York, NY 10010, USA}

\author[0000-0003-4511-8427]{Peter Kosec}
\affiliation{Center for Astrophysics | Harvard \& Smithsonian, Cambridge, MA, USA}

\author[0000-0003-4054-7978]{Riccardo Arcodia}
\affiliation{Black Hole Initiative at Harvard University, 20 Garden Street, Cambridge, MA 02138, USA}

\author[0000-0002-8304-1988]{Itai Linial}
\affiliation{Center for Cosmology and Particle Physics, Physics Department, New York University, New York, NY 10003, USA}

\author[0000-0001-5674-8403]{Olivia Aspegren}
\affiliation{Department of Astronomy \& Theoretical Astrophysics Center, University of California, Berkeley, CA 94720, USA}

\author[0000-0001-9735-4873]{Ehud Behar}
\affiliation{Department of Physics \& Kavli Institute for Astrophysics and Space Research, Massachusetts Institute of Technology, Cambridge, MA 02139, USA}
\affiliation{Department of Physics, Technion, Haifa 32000, Israel}

\author[0000-0002-6351-5808]{Sudip Bhattacharyya}
\affiliation{Department of Astronomy and Astrophysics, Tata Institute of Fundamental Research, 1 Homi Bhabha Road, Colaba, Mumbai 400005, India}

\author[0000-0002-1329-658X]{Margherita Giustini}
\affiliation{Centro de Astrobiolog\'ia (CAB), CSIC-INTA, Camino Bajo del Castillo s/n, 28692 Villanueva de la Ca\~nada, Madrid, Spain}

\author[0000-0002-8606-6961]{Lorena Hernández-García}
\affiliation{Instituto de Estudios Astrof\'isicos, Facultad de Ingenier\'ia y Ciencias, Universidad Diego Portales, Av. Ej\'ercito Libertador 441, Santiago, Chile}
\affiliation{Centro Interdisciplinario de Data Science, Facultad de Ingenier\'ia y Ciencias, Universidad Diego Portales, Av. Ej\'ercito Libertador 441, Santiago, Chile}

\author[0000-0002-5981-1022]{Daniel Kasen}
\affiliation{Department of Astronomy \& Theoretical Astrophysics Center, University of California, Berkeley, CA 94720, USA}
\affiliation{Department of Physics, University of California, Berkeley, CA 94720, USA}
\affiliation{Nuclear Science Division, Lawrence Berkeley National Laboratory, Berkeley, CA 94720, USA}

\author[0000-0003-0707-4531]{Giovanni Miniutti}
\affiliation{Centro de Astrobiolog\'ia (CAB), CSIC-INTA, Camino Bajo del Castillo s/n, 28692 Villanueva de la Ca\~nada, Madrid, Spain}

\author[0000-0001-9225-6481]{Frits Paerels}
\affiliation{Department of Physics and Columbia Astrophysics Laboratory, Columbia University, New York, NY 10027, USA}

\author[0000-0002-7116-2897]{Erwan Quintin}
\affiliation{European Space Agency (ESA), European Space Astronomy Centre (ESAC), Camino Bajo del Castillo s/n, 28692 Villanueva de la Cañada, Madrid, Spain}

\author[0000-0001-5231-2645]{Claudio Ricci}
\affiliation{Department of Astronomy, University of Geneva, ch. d’Ecogia 16, 1290, Versoix, Switzerland}

\author[0000-0002-5359-9497]{Daniele Rogantini}
\affiliation{Department of Astronomy and Astrophysics, University of Chicago, 5640 S Ellis Avenue, Chicago, IL 60637, USA}

\author[0000-0003-0820-4692]{Paula Sánchez-Sáez}
\affiliation{European Southern Observatory, Karl-Schwarzschild-Strasse 2, 85748 Garching bei München, Germany}

\author[0000-0003-0931-0868]{Fatima Zaidouni}
\affiliation{Department of Physics \& Kavli Institute for Astrophysics and Space Research, Massachusetts Institute of Technology, Cambridge, MA 02139, USA}

\begin{abstract}
Quasi-periodic eruptions (QPEs) are recurring bursts of X-ray radiation originating from supermassive black holes (SMBHs). They are an unprecedented type of structured, high-amplitude SMBH variability, but the physical origins of their regularity, timescales, energetics, and emission are uncertain. We present new \textit{XMM-Newton} observations of the QPEs in ZTF19acnskyy/``Ansky'', constituting the deepest observations of individual bursts in any source thus far. The X-ray spectra reveal time-evolving P\,Cygni profiles comprising blueshifted absorption and redshifted emission from L-shell transitions of Fe \textsc{xix}--\textsc{xxiv}, with column densities $N_H\sim 10^{22-23}$ cm$^{-2}$ and bulk velocities of $|v_w/c|\sim 0.2$, indicating relativistic mass ejections during each eruption. We construct a time-dependent analytical model of a wind turning on to self-consistently compute its evolving luminosity and ionization properties, and find that the light curve and spectral lines can be simultaneously produced by a wide-angle outflow with $\dot{M}\sim 10^{-9}-10^{-8}\,M_\odot$ s$^{-1}$ kinetically powering the X-rays with an efficiency of $L_X/\dot{E}_K\sim 0.1$. For a covering fraction $f_\Omega\sim0.5$, each eruption ejects $\sim 10^{-3}\,M_\odot$ and $\gtrsim 10^{49}$ erg of kinetic energy, setting an upper bound on the QPE lifetime of $\lesssim1000$ bursts if the underlying mass reservoir is $\sim1 M_\odot$, and implying that the bursts may result in detectable multiwavelength signatures of reverberation and feedback. These measurements provide new quantitative constraints on QPE energetics, emission mechanisms, and the mass/energy they recycle into their circumnuclear environments, as well as an observational probe for direct comparison with physical models and hydrodynamical simulations of QPEs.
\end{abstract}


\section{Introduction} \label{sec:intro}
Quasi-periodic eruptions (QPEs) are repeating soft X-ray flares observed from the nuclei of nearby low-mass galaxies \citep{Miniutti19,Giustini20,Arcodia21,Arcodia24a,Arcodia25,Chakraborty21,Chakraborty25a,Quintin23,Nicholl24,Hernandez25a,Baldini26}. The known population reaches peak luminosities of $L_{\rm peak}\sim10^{41-43}$ erg s$^{-1}$, recurs on timescales of $\sim$2.5--300 hr, and emits quasi-thermal spectra with $k_BT\sim50-250$ eV, implicating the innermost environments of supermassive black holes (SMBHs) with masses $\sim10^{5-7.5}\,M_\odot$. In several sources, the eruptions emerged months to years after a tidal disruption event (TDE), tying QPEs to their newborn accretion flows \citep{Chakraborty21,Chakraborty25a,Miniutti23a,Quintin23,Nicholl24,Baldini26}. Current estimates suggest $9^{+9}_{-5}\,\%$ of optically selected TDEs show QPEs \citep{Chakraborty25a}, comparable to independent estimates of their all-sky volumetric abundance of $0.60^{+4.73}_{-0.43}\times10^{-6}$ Mpc$^{-3}$ \citep{Arcodia24b}. For a more extensive review of their observational properties, see \cite{Wevers26}.

QPEs have generated interest as an unprecedented type of structured high-amplitude variability from SMBHs, which typically vary stochastically at the tens of percent level \citep[e.g.][]{Kara25}, and also as the possible counterparts of extreme mass-ratio inspirals \citep[EMRIs; e.g.][]{Xian21,Metzger22,Lu23,Franchini23,Linial23b,Tagawa23,Kejriwal24,Zhou24}. While EMRI models show promise in explaining many QPE observables---particularly their regularity, timescales, rates, and energetics---there remain uncertainties regarding their radiation mechanisms \citep{Vurm25,Huang25,Jankovic26}, formation channels and evolution \citep{Linial23a, Linial23b, Linial24c,Huang26,Yao26a}, mass and energy budgets \citep{Mummery25a,Linial25}, short- and long-term timing variations \citep{Miniutti25,Chakraborty25c,Chakraborty26a,Arcodia26}, and broader implications for dynamics in galactic nuclei \citep{Rom25}. Alternative theories model the bursts as instabilities or precession of the SMBH accretion disk \citep[e.g.][]{Raj21,Pan22,Sniegowska23,Middleton25}; while they have not yet been developed to the sophistication of EMRI-based models, making fair model comparison difficult, they remain a worthwhile topic for further investigation.

One relatively underexplored observational probe of QPEs is (time-resolved) X-ray spectroscopy, which has the potential to shed new light on the processes by which they generate emission and eject mass into their circumnuclear environments, with implications for collision dynamics and long-term fates of the systems. Thus far, high-resolution X-ray spectroscopy of the QPE host GSN 069 has detected persistent, steady-state outflows at $v\sim2000$ km s$^{-1}$ and $\dot{M}_{\rm out}\sim0.03-0.08\,M_\odot$ yr$^{-1}$ \citep{Kosec25}, while the source AT2019wzc/ZTF19acnskyy/``Ansky'' shows ionization features that change over individual bursts \citep{Chakraborty25b}. Both are consistent with super-solar nitrogen abundances \citep{Kosec25,Chakraborty25b,Wu26}, as expected if their long-term transient activities prior to the onset of QPEs were driven by tidal disruptions of evolved stars \citep[e.g.][]{Mockler24}. However, the outflows in the two sources likely have different origins; \cite{Kosec25} attribute the outflow in GSN 069 to a persistent wind from the long-term TDE evolution which formed the accretion disk rather than the QPEs themselves, supported by its lack of significant variability between or within eruptions. The more rapid and dramatic evolution in Ansky, as well as the significantly larger outflow velocities and broadening, implies that a different process drives its outflow, likely related to the QPE-emitting mechanism itself.

In this paper, we report new \textit{XMM-Newton} observations of Ansky, the brightest, longest-duration, and most energetic QPE source known: throughout 2025, its eruptions lasted $\sim$2 days, recurred every $\sim$10--13 days, and carried integrated energy budgets $\gtrsim100\times$ larger than those of typical QPEs \citep{Hernandez25a}. The eruptions began after an unusual long-lived optical/UV transient, possibly a slow featureless TDE produced by the disruption of an evolved star or a gradually fading low-luminosity AGN \citep{Sanchez24,Sanchez26,Zhu25}. Ansky has shown surprising timing behavior, with a secularly \textit{increasing} period \citep{Chakraborty26a,Hernandez25b} in contrast to most model predictions, and hosts the only multiwavelength counterpart to the X-ray bursts detected in any QPE source \citep{Guo26}. In this work, our focus is analysis and modeling of the broad, rapidly evolving absorption and emission lines in the X-ray spectra emerging during the QPEs, which we identify as a time-evolving P Cygni profile caused by mildly relativistic mass ejections powering the bursts.

In Section~\ref{sec:methods}, we describe the observations and methods used in our analysis. In Section~\ref{sec:results}, we present the results and a physical model to interpret the data. We discuss their implications in Section~\ref{sec:discussion} and conclude in Section~\ref{sec:conclusion}.

\section{Observations and methods} \label{sec:methods}

\begin{figure*}
    \centering
    \includegraphics[width=\textwidth]{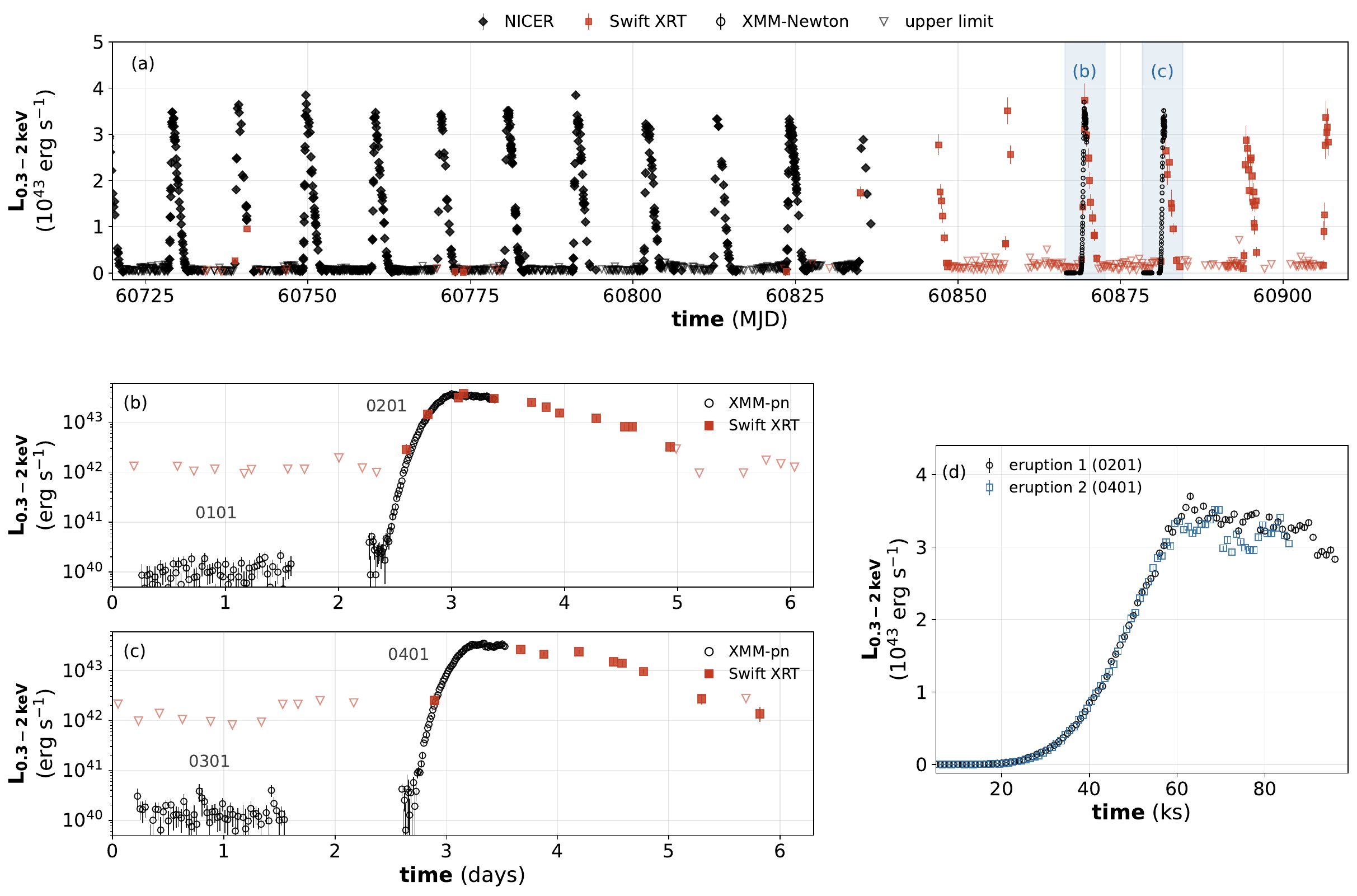}
    \caption{X-ray light curves of Ansky in 2025. \textbf{(a):} Long-term monitoring with \textit{NICER} (black diamonds), \textit{Swift} (red squares), and \textit{XMM-Newton} (open circles). Open triangles denote upper limits and shaded bands mark the two eruptions observed by \textit{XMM}, expanded in the lower panels. \textbf{(b)-(c):} Zoom-in of the \textit{XMM} observations of quiescence (OBSIDs 0101/0301) and eruptions (OBSIDs 0201/0401). \textbf{(d):} \textit{XMM} eruption light curves overlaid with their peaks aligned. The rise profiles are indistinguishable, while the eruptions diverge in their stochastic variability after the peaks.}
    \label{fig:lc}
\end{figure*}

\textit{XMM-Newton} data were obtained through Large Program 096454 in AO 24 (PI: Chakraborty). The observations were taken on July 10/12/22/24, 2025 (OBSIDs 0964540101-0964540401, hereafter 0101-0401), with the first and third (0101 and 0301) observing the quiescent state and the second and fourth (0201 and 0401) observing the rise-to-peak of two consecutive eruptions. The data were reduced using \textit{XMM} SASv21.0.0 and HEASoft v6.33. EPIC source products were extracted from a circular region of 33$''$ radius, while the background was extracted from a source-free circular region falling on the same detector with a 60$''$ radius. We retained events with PATTERN$\leq$4 (single and double events only) and discarded time intervals with a 10--12 keV count rate $\geq 1$ counts s$^{-1}$. Light curves were extracted in the 0.3--2 keV band with \texttt{evselect} and corrected for detector efficiency, vignetting, PSF, and bad pixels using \texttt{epiclccorr}. Despite the operation of the EPIC-pn camera in small window mode, pile-up remains a potential concern for Ansky's soft spectra reaching up to $\sim20$ cps. We thus checked whether extracting source products in an annular region, with varying inner radii excised between $5''/10''/15''$, affected the spectral products and found no significant changes among the different source regions. Moreover, similar spectral residuals are seen in both \textit{NICER} and RGS, neither of which is affected by pile-up. EPIC-pn count rates were converted to luminosities using a factor of $1\mathrm{\;cps}= 1.01\times10^{-12}$ erg cm$^{-2}$ s$^{-1}$, estimated for a 100 eV blackbody. RGS data were reduced with the \texttt{rgsproc} pipeline with a source extraction region centered on Ansky. Background flares were screened at $\geq0.25$ counts s$^{-1}$, and we used observational background spectra but verified our results against model backgrounds as well. RGS1 and RGS2 spectra were not combined, but fit separately with a free normalization that never exceeded $\sim$few percent. We binned the RGS spectra by a factor of 3 to mildly oversample the instrumental resolution. Spectral fitting was performed using \texttt{SPEX} \citep{Kaastra96} and \texttt{XSPEC} \citep{Arnaud96} with the Cash statistic \citep{Cash1979}.

The \textit{NICER} X-ray Timing Instrument \citep{Gendreau16} aboard the International Space Station observed Ansky for a total of 282.5~ks across 263 observations from Jan 5 to Jun 16, 2025 (PI: Chakraborty). We followed the time-resolved spectroscopy approach for light curve estimation outlined in Section 2.1 of \cite{Chakraborty24}. Spectral fitting and background estimation were performed with the \texttt{SCORPEON} model over a broadband energy range (0.25--10 keV) for data taken in orbit night, and a restricted range (0.38--10 keV) during orbit day. We grouped our spectra with the optimal binning scheme of \cite{Kaastra16} (\texttt{grouptype=optmin} with \texttt{groupscale=10} in the \texttt{ftgrouppha} command).

\textit{Swift} X-ray telescope (XRT) data were obtained from the online interface (\href{https://www.swift.ac.uk/LSXPS}{https://www.swift.ac.uk/LSXPS}) to the Living \textit{Swift} XRT Point Source (LSXPS) Catalogue \citep{Evans23a}, an automatically updated repository of all \textit{Swift} XRT observations of $>100$ second duration in Photon Counting (PC) mode. LSXPS reports X-ray fluxes in counts per second; we converted to cgs units using a conversion factor of $1\;\rm cps=3.1\times 10^{-11}$ erg cm$^{-2}$ s$^{-1}$ for a 100 eV blackbody.

To characterize rapid sub-eruption variability, we compute rms-normalized periodograms \citep{Vaughan03} of the 100-s binned, background-subtracted 0.3--2 keV EPIC-pn light curves, split at the light-curve peak into rise and post-maximum segments. Each periodogram is fit with a power law plus constant Poisson noise floor, $M(f)=N_0f^{-\alpha}+2\langle\sigma_{\rm err}^2\rangle\Delta t/\mu^2$, where $N_0$ is the power-law normalization, $\alpha$ the power-law index, $\Delta t=100$~s the time-bin width, $\mu$ the mean count rate of the segment, and $\langle\sigma_{\rm err}^2\rangle$ the mean squared count-rate uncertainty of the light-curve bins, so that the second term is the expected white-noise level in rms-normalized units. Because periodogram values are exponentially distributed about the true spectrum, we fit with the Whittle likelihood \citep[e.g.][]{Vaughan10}:
\begin{equation}
    \ln\mathcal{L}(N_0,\alpha) = -\sum_j\left[\frac{P_j}{M(f_j)} + \ln M(f_j)\right],
    \label{eq:whittle}
\end{equation}
where $P_j$ are the measured variability powers at frequencies $f_j$. The fractional variability is computed per posterior sample as $F_{\rm var}=[\int(M-C)\,df]^{1/2}$ over the 0.25--3 mHz band. We repeat the procedure in twelve energy bands to construct rms spectra. Power above the noise floor is considered detected when the likelihood ratio against the floor-only model exceeds $\Delta\ln\mathcal{L}=2$, which is equivalent to the noise-only model being excluded at one-sided 95\% confidence. Non-detections are thus quoted as 95\% upper limits, and energy bands with consecutive non-detections are combined into wider bands to increase sensitivity to low-amplitude variability.

\section{Results} \label{sec:results}

\begin{figure}
    \centering
    \includegraphics[width=\columnwidth]{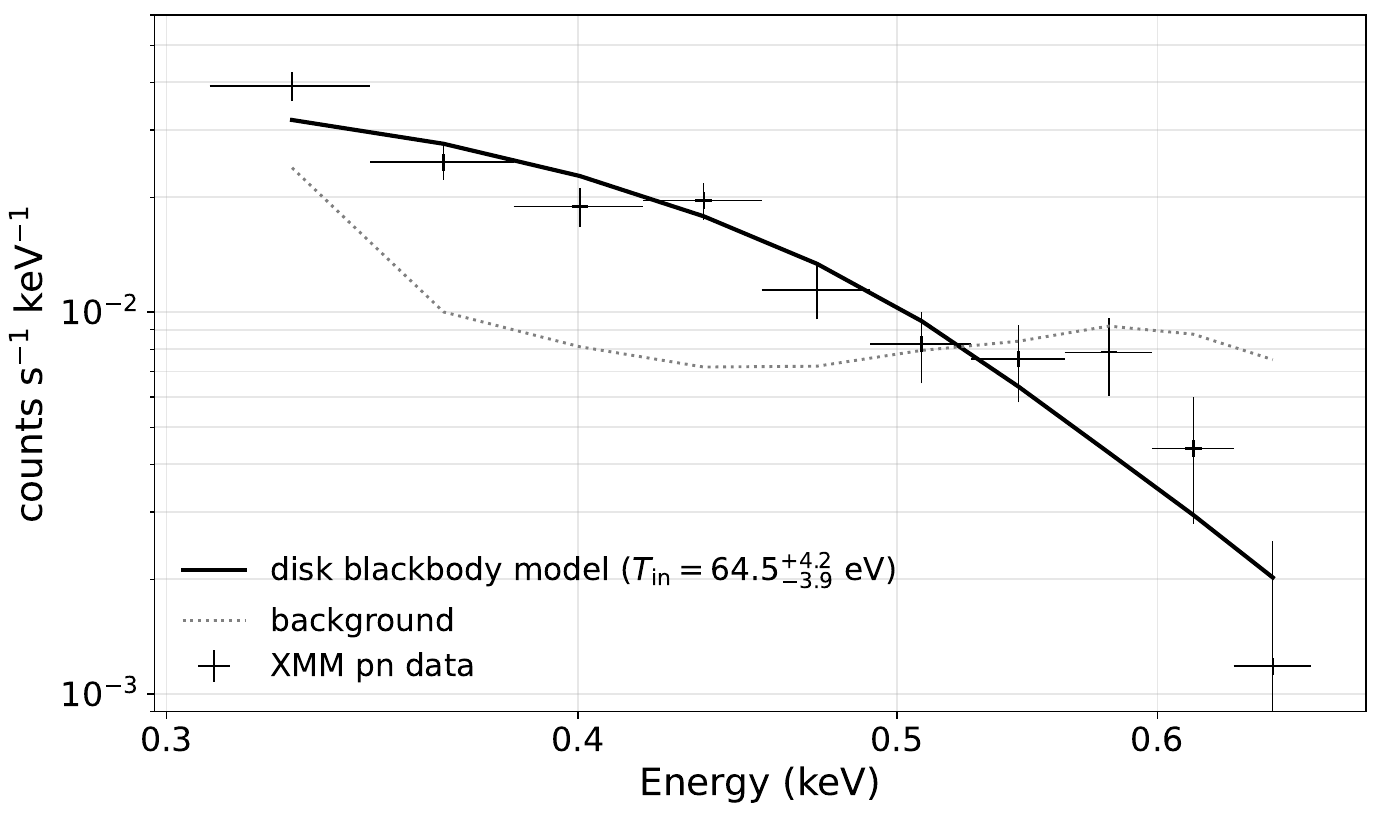}
    \caption{Stacked EPIC-pn spectrum of the quiescent state (OBSIDs 0101$+$0301; 164 ks total exposure), fit with a thermal disk blackbody model over 0.3--0.7 keV. The fitted model has $k_BT_{\rm in}=64^{+4}_{-4}$ eV and $L_{0.3-2}=1.9\times10^{40}$ erg s$^{-1}$.}
    \label{fig:quiescence}
\end{figure}

\begin{figure*}
    \centering
    \includegraphics[width=\textwidth]{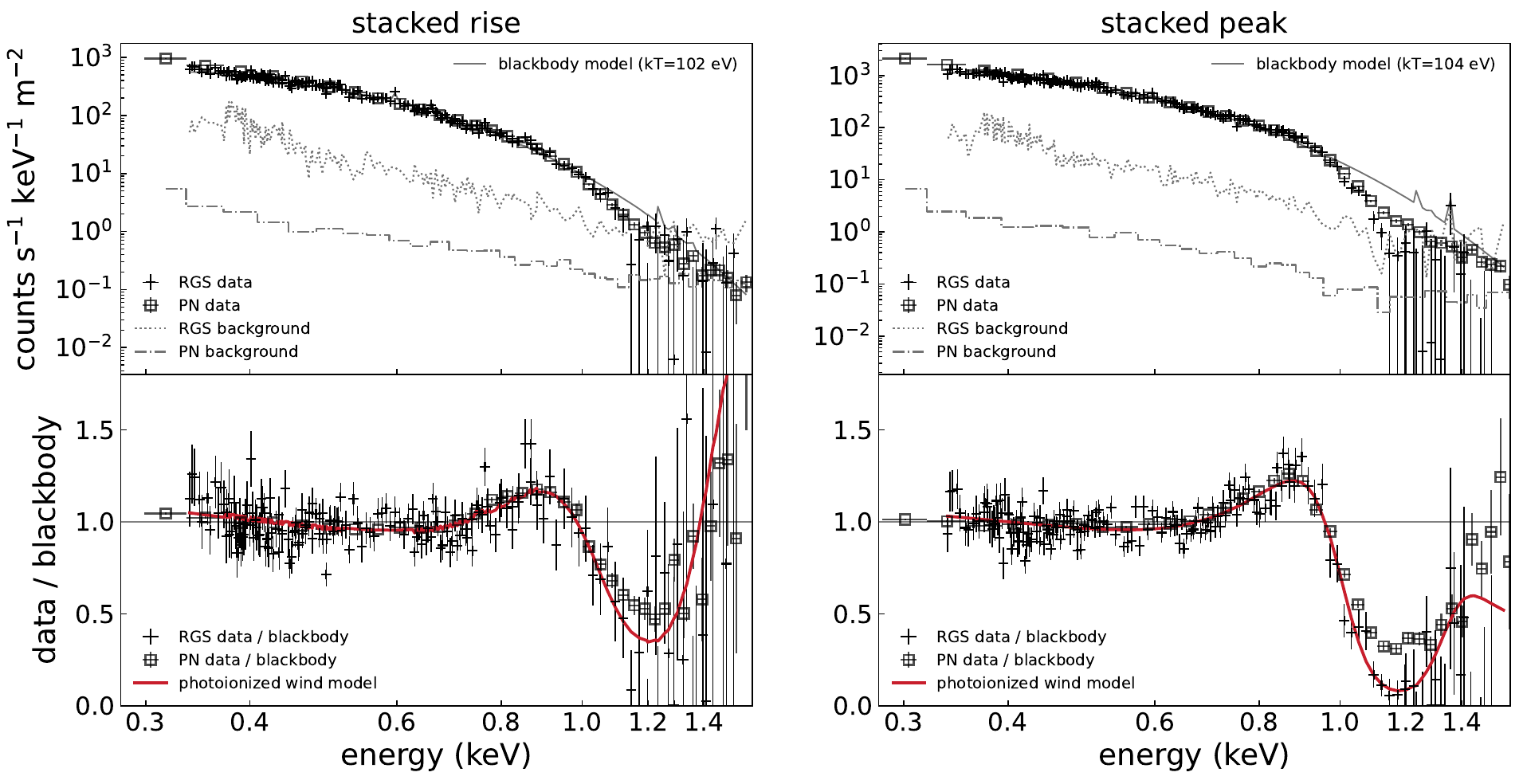}
    \caption{Simultaneous \textit{XMM-Newton} RGS and EPIC-pn spectra, stacked over the combined eruption rises (left) and peaks (right) of both observations. Upper panels show the data with single-blackbody continuum models. The best-fit blackbody temperatures $k_BT$ are annotated, and disagree by 3 eV between RGS and PN, likely due to instrument calibration differences. Lower panels show the data-to-model ratio, revealing the P Cygni profile near $\sim1$~keV that deepens from rise to peak.}
    \label{fig:rgs_pn}
\end{figure*}

\begin{table}
\centering
\caption{Best-fit parameters (with 1$\sigma$ uncertainties) of the two-component photoionized-wind model fit to the stacked rise and peak RGS spectra.}
\label{tab:pion}
\begin{tabular}{llc}
\toprule
Component & Parameter & Value \\
\midrule
\multicolumn{3}{c}{\textbf{Rise}} \\
\midrule
Blackbody                   & $k_BT$ (eV)                       & $94.0^{+2.2}_{-2.1}$ \\
\midrule
\multirow{4}{*}{Emission}   & $N_{\rm H}$ ($10^{22}$ cm$^{-2}$) & $3.0^{+2.5}_{-0.7}$ \\
                            & $\log\xi$ (erg cm s$^{-1}$)       & $3.63^{+0.99}_{-0.41}$ \\
                            & $v_{\rm turb}$ ($c$)              & $0.35^{+0.07}_{-0.10}$ \\
                            & $v_{\rm bulk}$ ($c$)              & $+0.11^{+0.08}_{-0.08}$ \\
\midrule
\multirow{4}{*}{Absorption} & $N_{\rm H}$ ($10^{22}$ cm$^{-2}$) & $22.4^{+8.1}_{-5.2}$ \\
                            & $\log\xi$ (erg cm s$^{-1}$)       & $6.54^{+0.25}_{-0.28}$ \\
                            & $v_{\rm turb}$ ($c$)              & $0.11^{+0.04}_{-0.01}$ \\
                            & $v_{\rm bulk}$ ($c$)              & $-0.13^{+0.07}_{-0.10}$ \\
\midrule
\multicolumn{2}{l}{$C$-stat/d.o.f.}                             & $1660/1470\approx1.13$ \\
\midrule
\multicolumn{3}{c}{\textbf{Peak}} \\
\midrule
Blackbody                   & $k_BT$ (eV)                       & $98.1^{+0.7}_{-1.8}$ \\
\midrule
\multirow{4}{*}{Emission}   & $N_{\rm H}$ ($10^{22}$ cm$^{-2}$) & $3.3^{+0.2}_{-0.7}$ \\
                            & $\log\xi$ (erg cm s$^{-1}$)       & $5.06^{+0.01}_{-0.70}$ \\
                            & $v_{\rm turb}$ ($c$)              & $0.25^{+0.03}_{-0.01}$ \\
                            & $v_{\rm bulk}$ ($c$)              & $+0.23^{+0.01}_{-0.23}$ \\
\midrule
\multirow{4}{*}{Absorption} & $N_{\rm H}$ ($10^{22}$ cm$^{-2}$) & $9.8^{+1.3}_{-1.2}$ \\
                            & $\log\xi$ (erg cm s$^{-1}$)       & $6.32^{+0.21}_{-0.09}$ \\
                            & $v_{\rm turb}$ ($c$)              & $0.08^{+0.01}_{-0.01}$ \\
                            & $v_{\rm bulk}$ ($c$)              & $-0.17^{+0.04}_{-0.08}$ \\
\midrule
\multicolumn{2}{l}{$C$-stat/d.o.f.}                             & $1990/1464\approx1.36$ \\
\bottomrule
\end{tabular}
\end{table}

We show the 2025 X-ray light curve of Ansky in Fig.~\ref{fig:lc}, combining our \textit{NICER} and \textit{Swift} monitoring with the four \textit{XMM} observations described in Section~\ref{sec:methods}. OBSIDs 0101 and 0301 observed only the quiescent state, which emits a thermal disk blackbody spectrum with $L_{q,0.3-2}\approx1.9\times10^{40}$ erg s$^{-1}$ and an inner edge temperature $k_BT_{\rm in}=64^{+4}_{-4}$ eV; in Fig.~\ref{fig:quiescence}, we show the data fitted by the \texttt{diskbb} model in \texttt{XSPEC}. OBSIDs 0201 and 0401 caught the rise and peak of two consecutive eruptions separated by 12.1 d. The eruptions climb by a factor of $\sim 1500$ to $L_{\rm pk,0.3-2}\approx 2.7\times10^{43}$ erg s$^{-1}$ over 30 ks, while the \textit{Swift} monitoring tracks their subsequent slow decline back to quiescence.

The two eruption rises are nearly identical. In panel (d) of Fig.~\ref{fig:lc} we overlay the two \textit{XMM} light curves with their peaks aligned; the rise profiles agree at 1-ks resolution over the full $\gtrsim3$ decades in flux, and only decorrelate in the stochastic flickering after the peak. This near-perfect similarity in amplitude, duration, spectral evolution, and profile shape demonstrates remarkably reproducible eruption dynamics and launch conditions, suggesting a well-regulated central engine within the observed window.

\subsection{P Cygni profile during the eruptions} \label{subsec:pcyg}

In Fig.~\ref{fig:rgs_pn} we show simultaneous RGS and EPIC-pn spectra stacked over the combined rises and the combined peaks of both eruptions. They are well-described by the Wien tails of thermal blackbodies that heat to $\approx100$ eV at peak. The data-to-model ratios reveal two features imprinted on this thermal continuum: a broad, blueshifted absorption trough above $\sim$1 keV that deepens from rise to peak, and an excess of emission just redward of the trough, near $\sim$0.9 keV. This combination of blueshifted absorption and redshifted emission is the classic P\,Cygni signature of an outflow, in which the absorption arises from optically thick plasma directly occluding the continuum photosphere, and the emission from optically thin plasma in the rest of the outflow \citep[e.g.][]{CastorLamers79}. Such broadened and blueshifted outflows are sometimes seen in accreting SMBHs, such as AGN \citep[e.g.][]{Tombesi12,Masterson22,Xrism25} and TDEs \citep[e.g.][]{Kara18,Yao24}, though they do not typically vary on such short timescales.

To characterize the profile, we fit the stacked RGS spectra with a two-component photoionized plasma of solar chemical abundance, characterized by the \texttt{PION} spectral model \citep{Mehdipour16}. We include one component contributing only absorption lines, and another only in emission, reporting the results in Table~\ref{tab:pion}. The absorber's covering fraction was fixed at unity and the emitter's solid angle at $\Omega=4\pi$, so the emission column should be read as $f_\Omega N_{H,\rm em}$. Our fits show the absorber is a high-column density, highly ionized wind ($N_{\rm H}\approx10^{23}$ cm$^{-2}$, $\log\xi\approx6.3-6.5$) outflowing at $v_{\rm bulk}\approx-(0.1$--$0.2)c$, while the emission component is redshifted by a comparable magnitude, which can arise from the receding hemisphere of the flow in a biconical or asymmetric geometry. The velocities are mildly relativistic and the turbulent widths are comparable to $v_{\rm bulk}$, consistent with emission integrated over a wide range of projected velocities in the wind. However, we note that $v_{\rm turb}$ in these time-averaged spectra is overestimated due to blending of the moving absorption/emission profiles into one smeared profile; the $v_{\rm turb}$ values we obtain when fitting the time-resolved spectra are substantially lower (Section~\ref{subsec:physical_model}). The ionization parameters appear significantly higher than in the lower-luminosity 2024 \textit{XMM} observations, which were consistent with a range $\log\xi\sim1.9-4.2$ \citep{Chakraborty25b}. The assumption of solar abundances is approximate given the super-solar nitrogen found previously \citep{Kosec25,Chakraborty25b}; however, nitrogen is fully ionized at these ionization parameters and only Fe contributes to the observed lines, so thawing the Fe abundance would be exactly degenerate with $N_H$, changing it by at most a factor of a few.

\begin{figure}
    \centering
    \includegraphics[width=\columnwidth]{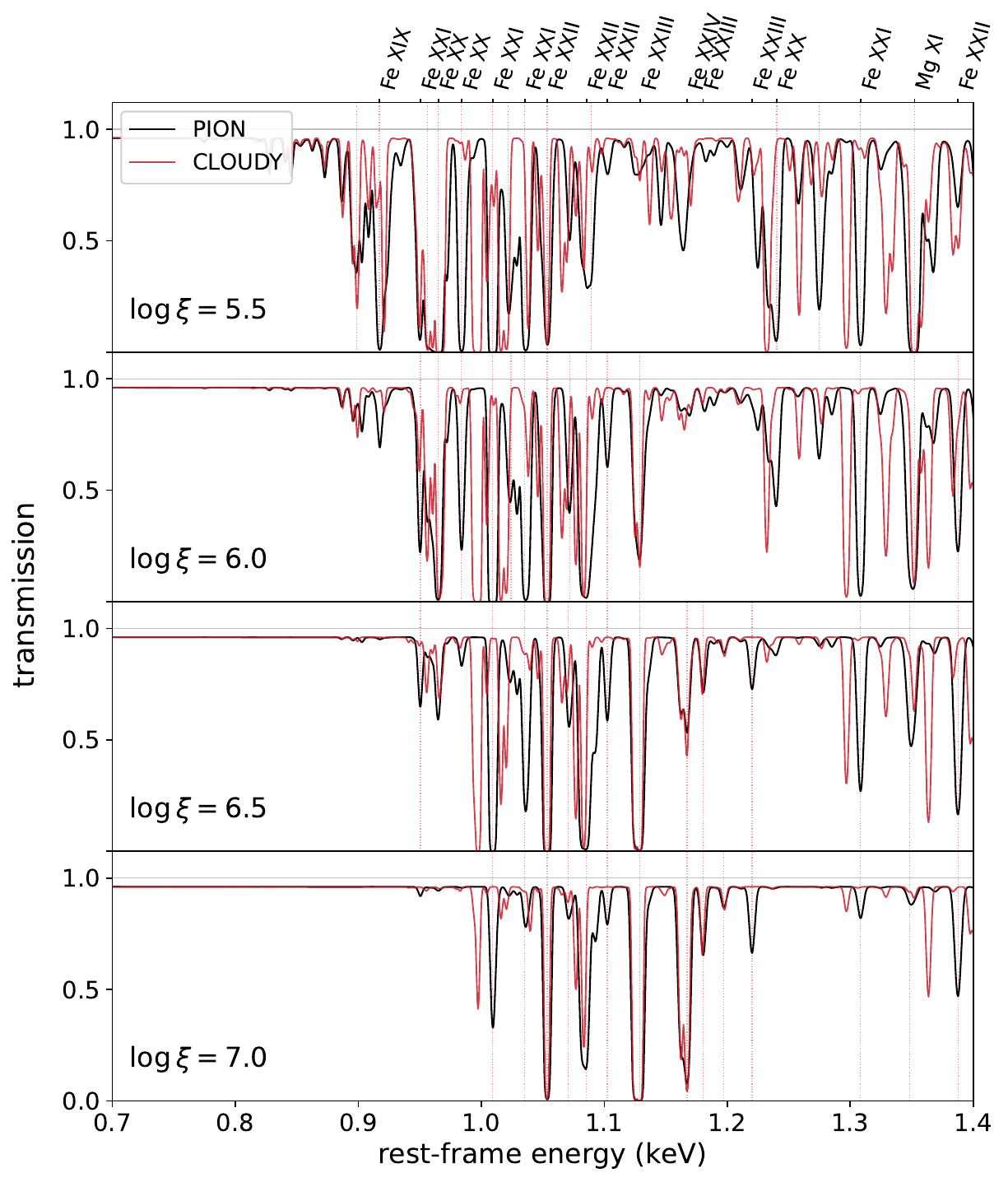}
    \caption{Rest-frame transmission spectra of photoionized plasma with $N_{\rm H}=5\times10^{22}$ cm$^{-2}$ and $v_{\rm turb}=500$ km s$^{-1}$ irradiated by a 100 eV blackbody continuum. Increasing $\xi$ shifts the lines to higher energies, which, for a broad unresolved line blend, can mimic the effect of a higher blueshift.
    }
    \label{fig:transmission_grid}
\end{figure}

The most remarkable feature of the P\,Cygni profile in Ansky is that it evolves in centroid, depth, and width on kilosecond timescales, as previously discussed in \cite{Chakraborty25b}. In that work, we modeled the evolution as being driven by the dynamically changing velocity, density, and ionization profiles of the ejecta in a homologous expansion. However, this did not account self-consistently for ionization effects as the wind responds to the changing continuum, which can mimic the effect of a changing outflow velocity as $v_w$ and the ionization parameter $\xi\equiv L_{\rm ion}/(nr^2)$ compete in setting the line centroids. We illustrate this in Fig.~\ref{fig:transmission_grid} with transmission spectra of fixed plasma slabs of column density $N_H=5\times10^{22}$ cm$^{-2}$ irradiated by blackbody spectra of $k_BT=100$ eV for varying $\log\xi\in\{5.5,6,6.5,7\}$ erg cm s$^{-1}$, computed using both \texttt{PION} and \texttt{CLOUDY} \citep{Ferland17} for independent estimates. The dominant absorption lines arise from $n=2\rightarrow3$ L-shell resonance transitions of highly ionized iron: $2p\rightarrow3d$ in Fe \textsc{xix}--\textsc{xxii} and $2s\rightarrow3p$ in the Be- and Li-like charge states Fe \textsc{xxiii}--\textsc{xxiv}. The weaker features above $\sim$1.2 keV correspond to the $n=2\rightarrow4$ members of the same ions. As $\xi$ increases, so does the ionization state of the absorbing plasma, resulting in a bulk shift of the line blend towards higher energies.

\begin{figure*}
    \centering
    \includegraphics[width=\textwidth]{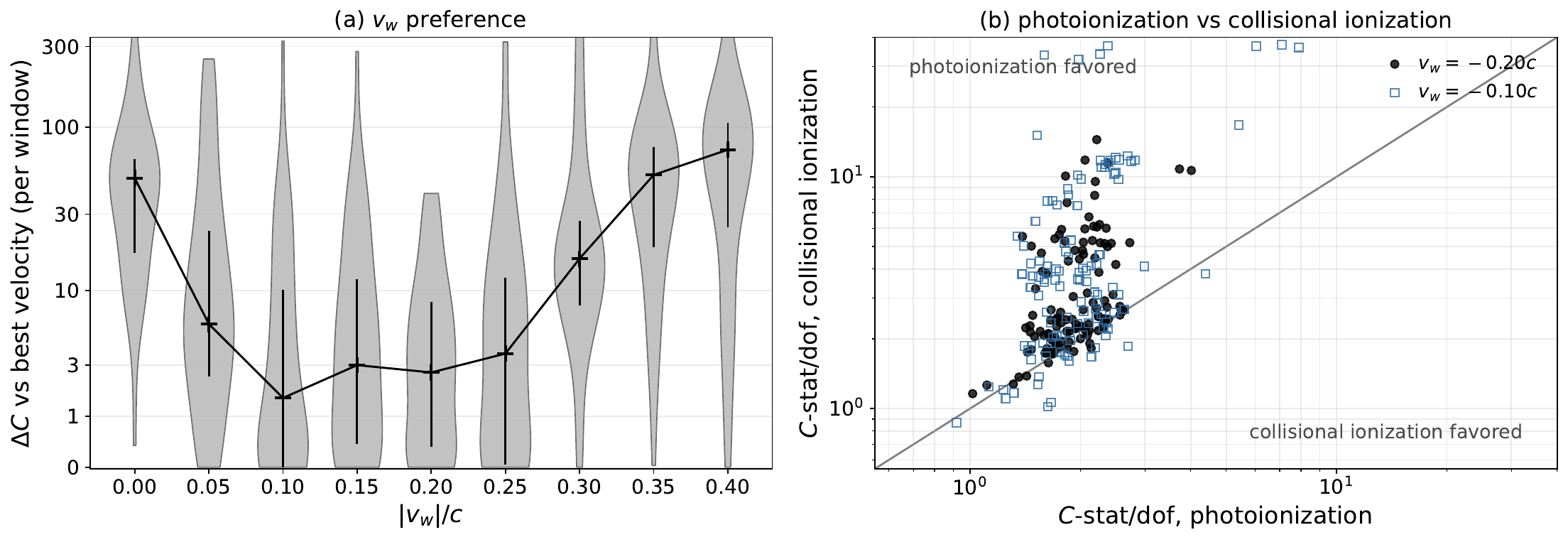}
    \caption{Model preference in the time-resolved fits. \textbf{(a):} Outflow velocity preference: for each window, $\Delta C(v_w)=C(v_w)-\min C(v)$ measures how much worse each $v_w$ is than the best one. The minimum is flat over $0.1c-0.2c$ and rises outside it, excluding both a static absorber and $v_w\gtrsim0.25c$. \textbf{(b):} Reduced fit statistics comparing collisionally ionized to photoionized absorption. In nearly all windows the collisional ionization model is disfavored.}
    \label{fig:model_pref}
\end{figure*}

To address the ionization/velocity degeneracy, we performed a grid of time-resolved spectral fits. We divided the \textit{XMM} EPIC spectra from both eruptions into time-resolved bins spanning 10~ks with a 1~ks timestep between windows. Due to the low SNR of the early low-flux windows, we stack up to five consecutive spectra (maximum 15~ks) to increase signal. We then fit each time-resolved spectrum with the same photoionized plasma emission+absorption model as the RGS data. Fitting the spectra with all parameters free is highly degenerate, so we iteratively fixed the outflow speed to determine the response of the other parameters. We tested nine different assumed speeds, $v_w/c=\{0,0.05,0.1,...,0.4\}$, holding them constant while leaving the ionization parameter, column density, and turbulent velocity broadening free to vary in each window to accommodate the changing P Cygni profile, and fit the emission component with a Gaussian. We quantify the model preference in Fig.~\ref{fig:model_pref}a, using the paired statistic $\Delta C(v_w)=C(v_w)-\min_{v}C(v)$, i.e.\ how much worse each choice of $v_w$ fits than the best one for a given window. Wind speeds of $v_w\lesssim0.1c$ or $\gtrsim 0.25c$ are disfavored, but the data are about equally well-fit by the models in the range $v_w=0.1-0.2c$. We discuss the full evolution of the resulting fit parameters for $v_w=\{0.1c,0.2c\}$ in Appendix~\ref{app:supplemental_figures}.

We also considered whether the outflow may be collisionally ionized rather than photoionized. To assess this, we fit each window in both \textit{XMM} observations with collisionally ionized absorption plasmas using the \texttt{HOT} model in \texttt{SPEX}, for assumed fixed outflow velocities $v_w/c\in\{0.1,0.2\}$ and a Gaussian component for the emission. We show the comparison of fit statistics in Fig.~\ref{fig:model_pref}b. The collisional model is worse in 85\% of the 234 windows, by a median $\Delta C=53$ and by $\Delta C>5$ in 78\% of them, indicating the dominant excitation process in the outflow is photoionization. We further discuss the fit residuals for a collisional plasma in Appendix~\ref{app:supplemental_figures}.

\subsection{A time-dependent wind model for the eruptions} \label{subsec:physical_model}

\begin{figure*}
    \centering
    \includegraphics[width=\textwidth]{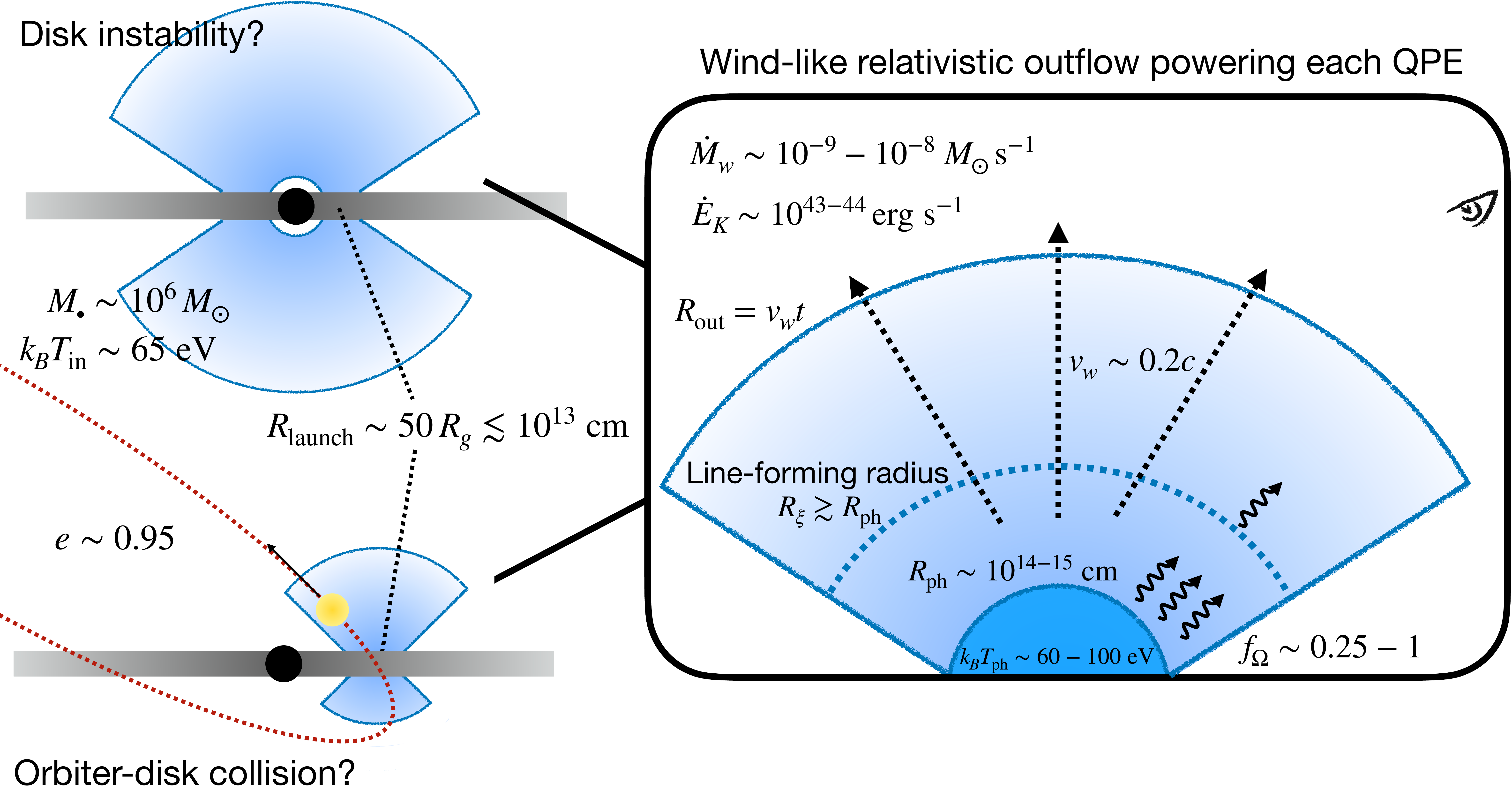}
    \caption{Schematic illustration (not to scale) of the eruption geometry implied by our measurements. The system has a persistent accretion disk with $k_BT_{\rm in}\approx65$ eV (Fig.~\ref{fig:quiescence}). During eruptions, material is launched close to the escape velocity at $R_{\rm launch}\sim50R_g$---e.g. by super-Eddington winds from disk instabilities or from matter expelled by orbiter-disk collisions in an EMRI---and expands as a wide-angle ($f_\Omega\sim 0.25$--$1$) mildly relativistic ($v_w\sim 0.2c$) wind-like outflow carrying $\dot M_w\sim10^{-9}-10^{-8}\,M_\odot$ s$^{-1}$ with a kinetic power $\dot E_K\sim10^{43-44}$ erg s$^{-1}$. The observed quasi-thermal continuum ($k_BT_{\rm ph}\sim 60-100$ eV) emerges from the photosphere, a region of size $R_{\rm ph}\sim10^{14-15}$ cm, and is imprinted with the blueshifted absorption features by the wind close to the line-forming radius at $R_\xi\gtrsim R_{\rm ph}$. The presence of redshifted emission lines implies a biconical or asymmetric outflow geometry.}
    \label{fig:schematic}
\end{figure*}

While independently fitting each spectrum is useful to assess the characteristic outflow parameters, it does not provide an \textit{explanation} for the evolution, as it is not a manifestly time-dependent model. We thus turn our attention to a physical forward model for the time-evolution of the spectral lines during the eruptions, depicted schematically in Fig.~\ref{fig:schematic}. Consider a wind launched at the eruption onset $t_0$, subtending a covering fraction $f_\Omega$ and coasting at a constant speed $v_w$, which powers a luminosity:
\begin{align}
    L_X(t) &= \epsilon_X \dot{E}_K=\tfrac{1}{2}\,\epsilon_X\,\dot M_w(t)\,v_w^2 \\
    \dot M_w(t) &= \frac{2L_X(t)}{\epsilon_X v_w^2}.
    \label{eq:mdot_model}
\end{align}
where $\dot{E}_K(t)$ is the kinetic power of the wind with mass outflow rate $\dot{M}_w(t)$. By assumption, we fix the radiative efficiency $\epsilon_X$ in time (though see \citealt{Linial25}). We also implicitly assume that the X-ray emission is produced at the outflow launching time; in reality, there may be a delay between the two, as the X-ray photons are initially trapped by the outflow. The photon trapping radius is given by $r_{\rm tr} = \kappa \dot{M}/(4\pi c) \sim 5\times10^{13}\mathrm{\,cm}$ for $\dot{M}=10^{-9}M_\odot\;\mathrm{s}^{-1}$ and $\kappa_{\rm eff}\approx8.5$ cm$^2$ g$^{-1}$, the effective opacity implied by our fits (Section~\ref{subsec:physical_model}). This gives an expansion timescale of $r_{\rm tr}/v_w \sim 8$ ks for $v_w = 0.2c$. Since this is much shorter than the rise timescale of the QPEs, it is appropriate to ignore this time delay for the simple model presented here.

To calculate the column density and ionization balance of atomic species in the wind --- which set the observed line profiles via the scattering depth and the balance between atomic excitation (via photoionization) and de-excitation processes (e.g. radiative recombination) --- one needs the radial density structure at each time in the evolving wind. In the simple case of a steady-state wind, the density profile is obtained from the equation of mass continuity, $4\pi f_\Omega r^2\rho(r,t)v_w = \dot M_w$. However, the situation is more complicated in our clearly non-steady-state case. After each radial mass shell is launched, causality requires that density at radius $r$ reflects the mass-loss rate at the retarded time:
\begin{equation}
    \rho(r,t) = \frac{\dot M_w\!\left(t - r/v_w\right)}{4\pi f_\Omega r^2 v_w} = \frac{L_X\!\left(t - r/v_w\right)}{2\pi f_\Omega \epsilon_X r^2 v_w^3}
    \label{eq:rho_retarded}
\end{equation}
truncated at outer edge $R_{\rm out}=v_w(t-t_0)$ beyond which the wind has not yet reached. Eq.~\ref{eq:rho_retarded} expresses the fact that the radial mass shell contributing to the absorption at time $t$ was the \textit{same} mass shell powering the eruption at time $t-r/v_w$.

The continuum photosphere is the surface above which the optical depth through the wind is of order unity. For a gray effective opacity $\kappa$, the optical depth is proportional to the overlying column,
\begin{equation}
    \tau(r,t)=\kappa\,\mu m_p\int_r^{R_{\rm out}(t)} n(r',t)\,dr',\qquad \tau(R_{\rm ph},t)\equiv1,
    \label{eq:tau_rph}
\end{equation}
where $\rho=\mu m_p n$ with $\mu m_p$ the mean molecular weight per hydrogen ion. The photospheric condition thus requires the column density above the photosphere to be constant,
\begin{equation}
    N_H \equiv \int_{R_{\rm ph}(t)}^{R_{\rm out}(t)} n(r,t)\,dr = \frac{1}{\kappa\,\mu m_p},
    \label{eq:nh_sat}
\end{equation}
independent of time and of the density profile.

Finally, the local ionization parameter, which sets the observable atomic transitions, is:
\begin{equation}
    \xi(r,t)\equiv\frac{L_X(t)}{n(r,t)\,r^2}
    = 2\pi f_\Omega\,\mu m_p\,\epsilon_X\,v_w^3\,\frac{L_X(t)}{L_X(t-r/v_w)}.
    \label{eq:xiss}
\end{equation}
The ionization parameter thus evolves over time and also spatially throughout the wind, requiring us to specify the location of the line-forming radius $R_\xi>R_{\rm ph}$ in the model. In total the time-dependent wind model thus has five free parameters: $\epsilon_X$, $v_w$, $f_\Omega$, $N_H$, and $\eta\equiv R_\xi/R_{\rm ph}$. We also note that $\epsilon_X$ and $f_\Omega$ are observationally degenerate, as they enter only as the product $\xi\propto f_\Omega\,\epsilon_X$ (Eq.~\ref{eq:xiss}); we discuss the implications of this at the end of this section.

We now compare this forward model to the time-resolved \textit{XMM} spectra described in Section~\ref{subsec:pcyg}. We first parametrize the luminosity evolution entering Eqs.~\ref{eq:mdot_model}--\ref{eq:xiss} as a power-law rise followed by a power law with an exponential cutoff:
\begin{equation}
    L_X(t) = L_q + L_{\rm pk}
    \begin{cases}
        \Big(\frac{t-t_0}{t_{\rm pk}-t_0}\Big)^{\alpha}, & t\le t_{\rm pk}\\[2pt]
        \Big(\frac{t-t_0}{t_{\rm pk}-t_0}\Big)^{\beta}\,e^{-(t-t_{\rm pk})/\tau_2}, & t>t_{\rm pk}
    \end{cases}
    \label{eq:plaw_g}
\end{equation}
We fit this functional form to the 0.3--2 keV luminosities of the time-resolved spectra together with the \textit{Swift} late-time data (Section~\ref{sec:methods}). For the first eruption this gives $L_{\rm pk}=3.1\times10^{43}$ erg s$^{-1}$, $t_{\rm pk}-t_0=43$ ks, $\alpha=3.0$, $\beta=1.6$, and $\tau_2=37$ ks, with 0.03 dex rms. For the second eruption we find $L_{\rm pk}=2.8\times10^{43}$ erg s$^{-1}$, $t_{\rm pk}-t_0=39$ ks, $\alpha=2.5$, $\beta=1.9$, and $\tau_2=35$ ks, with 0.04 dex rms.

With $L_X(t)$ in hand, we can now specify the rest of the model to compute $\xi(t)$ and $N_H(t)$ and determine whether the data can be reproduced. We compute the line profile evolution for each combination of parameters in $\epsilon_X\in\{0.01,0.1,1\}$, $v_w\in\{0.1c,\,0.2c\}$, $f_\Omega\in\{0.5,1\}$, $N_H\in\{10^{22},\,5\times10^{22},\,10^{23}\}$ cm$^{-2}$, and $R_\xi\in\{1.1,\,1.2,\,1.3\}\,R_{\rm ph}$, freezing $\xi$ and $N_{H,\rm abs}$ in each spectrum to the model values, with the emission component's ionization tied to the same values and its velocity mirrored to $+|v_w|$. In each time-resolved spectrum, this leaves four free parameters: the turbulent velocities of the emission and absorption plasmas, the continuum temperature, and the normalization of the emission component (which scales as $f_\Omega N_{H,\rm em}$, and need not share the same column as the absorber due to the P Cygni geometry). The total degrees of freedom are the sum of the spectral bins in the EPIC-pn/MOS1/MOS2 spectra across all $\sim50-60$ windows of each observation.

\begin{figure}
    \centering
    \includegraphics[width=1.1\columnwidth]{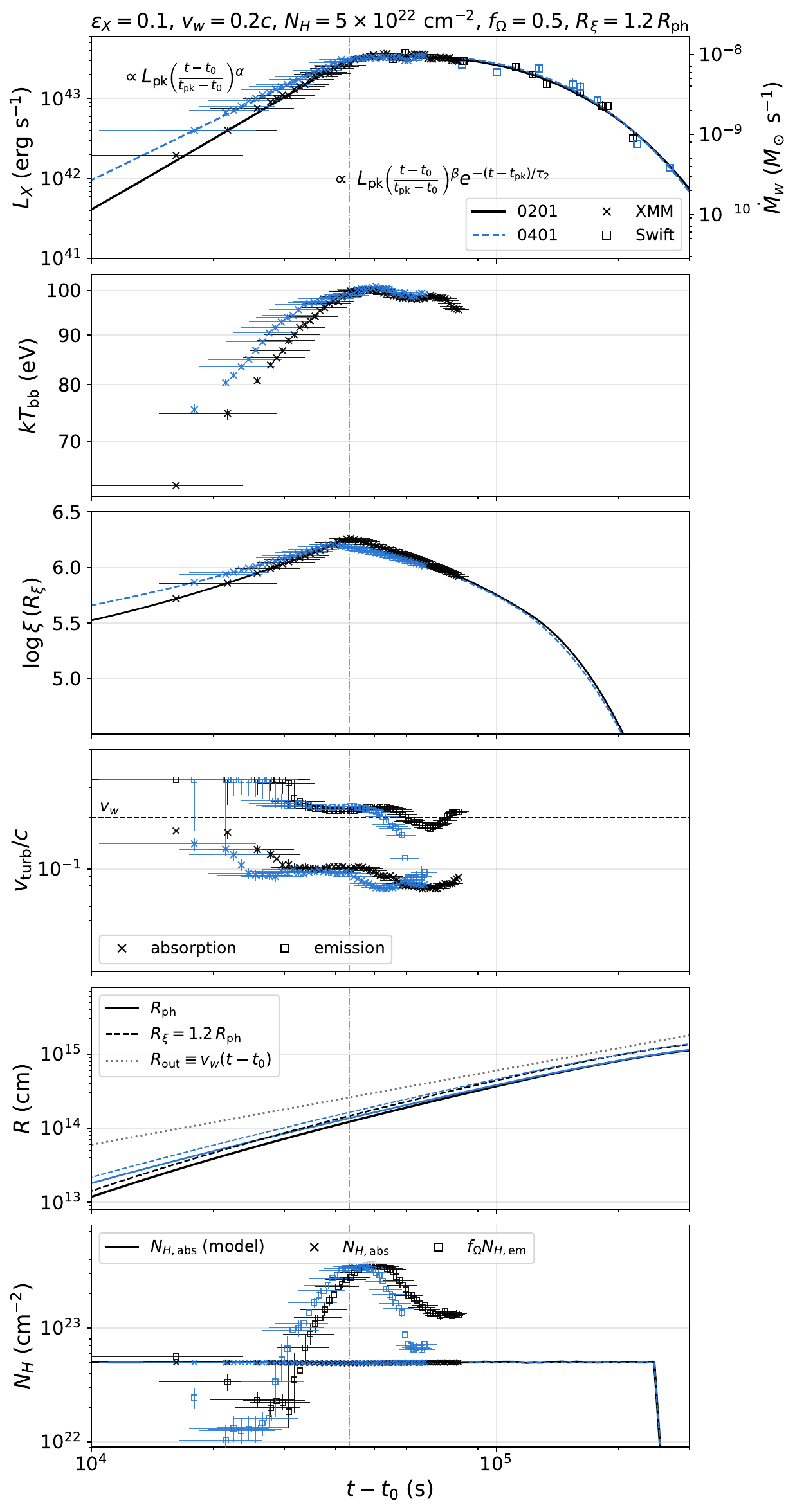}
    \caption{Time evolution of the wind model (Section~\ref{subsec:physical_model}) providing the best fit to our data $(\epsilon_X=0.1, v_w=0.2c, f_\Omega=0.5, N_H=5\times10^{22}, R_\xi=1.2R_{\rm ph})$. Data are shown for both eruptions (OBSID 0201/0401 in black/blue). From top to bottom, we show time evolution of luminosity, photosphere temperature, ionization parameter at the line-forming radius, turbulent velocities of the absorption/emission components, radial scales in the model, and column densities of the absorption/emission components.}
    \label{fig:model_evol}
\end{figure}

\begin{figure}
    \centering
    % \begin{interactive}{animation}{spec_evol.mp4}
    \includegraphics[width=\columnwidth]{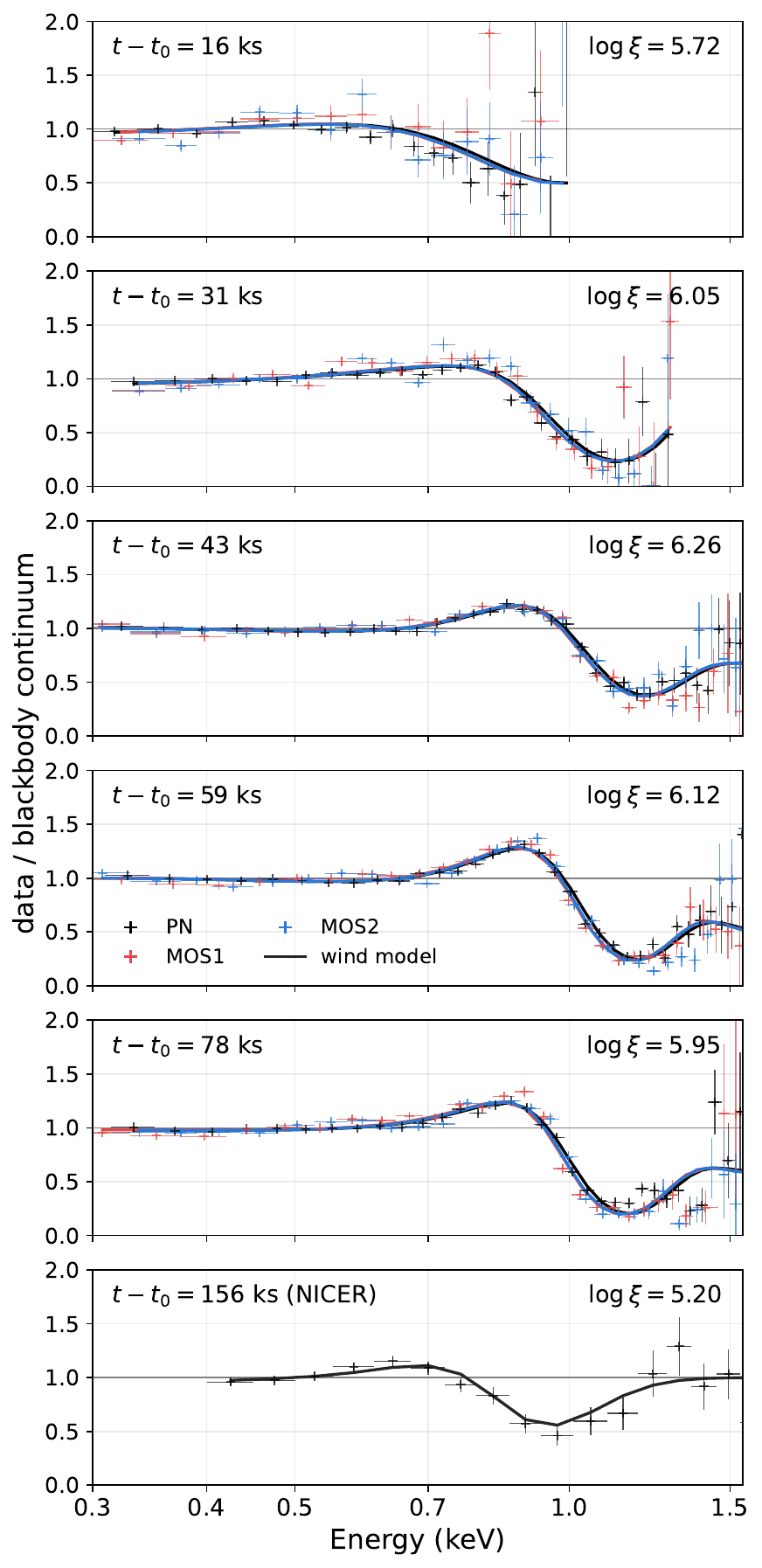}
    % \end{interactive}
    \caption{Time-resolved spectral fits with the wind model of Section~\ref{subsec:physical_model} for six representative spectra. Each panel shows the residuals to a blackbody continuum, with lines indicating the model emission/absorption throughout the eruption. The absorbing plasma is frozen at the wind model's predicted $(\log\xi,\,N_{H,\rm abs})$ at all times, while the emission component has frozen $\log\xi$ and free $f_\Omega N_{H,\rm em}$. The bottom panel is a late-time \textit{NICER} spectrum from a different eruption. An animation showing the full spectral evolution is available in the HTML version of the article, or at \href{https://joheen.com/qpe_spectral_evolution.gif}{https://joheen.com/qpe\_spectral\_evolution.gif}.}
    % The animation spans 7.6 s and comprises two three-panel columns simultaneously showing the light curve, spectral, and line evolution annotated with $t-t_0$, $\log\xi(t)$, and $N_H(t)$. It also shows a static strip of four late-time \textit{NICER} spectra.}
    \label{fig:ratio_evol}
\end{figure}

\begin{figure}
    \centering
    \includegraphics[width=\columnwidth]{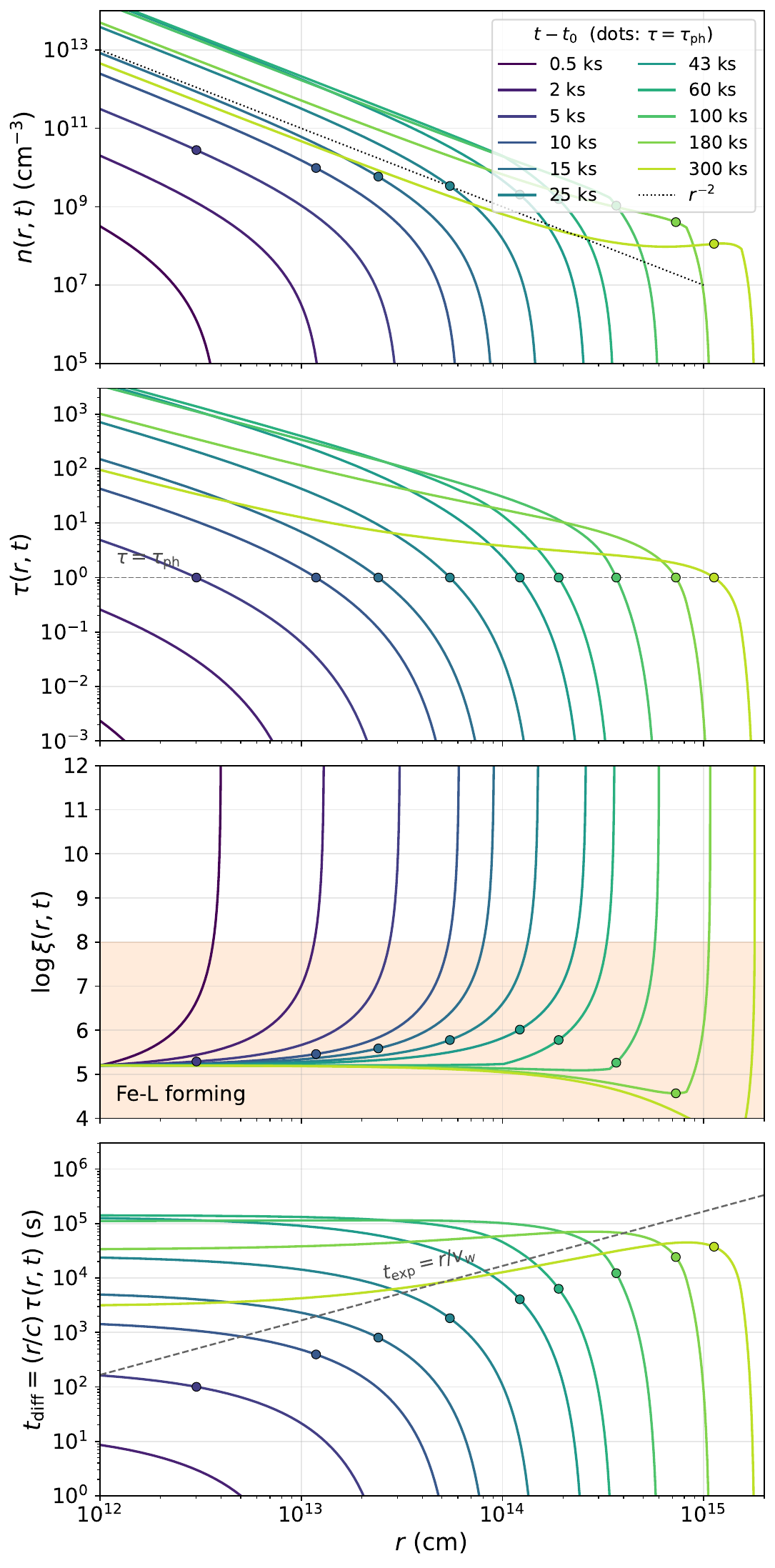}
    \caption{Radial structure of the wind model throughout an eruption. From top to bottom, we show the time-evolving profiles of ion number density, optical depth, ionization parameter, and local diffusion time in the wind. The filled circles in each panel mark the photosphere radii, where the optical depth integrates to one. The shaded region in the third panel marks the range of ionization parameters which can form the observed Fe-L absorption features.}
    \label{fig:wind_profiles}
\end{figure}

The data prefer $v_w=0.2c$, $N_H=5\times10^{22}$ cm$^{-2}$, $R_\xi=1.2R_{\rm ph}$, and $\epsilon_X f_\Omega=0.05$ (evaluated at $\epsilon_X=0.1$, $f_\Omega=0.5$). The best-fitting model has $C$-stat/dof $=9331/4656\approx 2.0$ for OBSID 0201 and $C$-stat/dof$=6621/3799\approx 1.7$ for OBSID 0401. We show the time evolution of continuum temperature, model ionization parameter, turbulent velocities, radial scales, and column densities for both eruptions in Fig.~\ref{fig:model_evol}. We show six representative time-resolved spectra in Fig.~\ref{fig:ratio_evol}; the last was taken from a late-time \textit{NICER} observation of a different burst on MJD 60710.7, which was not included in the fit, but demonstrates that the data indeed show a continued decrease in the line centroid as the model predicts. We also illustrate the radial structure of the wind---$n(r,t)$, $\tau(r,t)$, $\xi(r,t)$, and the diffusion time through the wind, $t_{\rm diff}(r,t)=\tau r/c$---at various times throughout an eruption in Fig.~\ref{fig:wind_profiles}. During the rise, the wind is steeper than the steady $r^{-2}$ profile (dotted), because gas at larger radii was launched when $\dot M_w$ was lower. The profile then approaches the steady-state $r^{-2}$ profile near peak and flattens outward in the decline. The column density $N_H\approx5\times10^{22}$ cm$^{-2}$ implies the product of the effective opacity and mean molecular weight is $\kappa\mu\approx 12$ cm$^2$ g$^{-1}$, $\approx25\times$ larger than the naive expectation using electron-scattering opacity ($\kappa_{\rm es}\approx0.34$ cm$^{2}$ g$^{-1}$) and $\mu=1.4$ for solar abundance plasma. This reflects that the bound-bound and bound-free transitions contribute significantly to the soft X-ray opacity.

Most remaining regions of the parameter grid are strongly disfavored. Models with $v_w=0.1c$ all have $\Delta C\geq+503$ relative to the best model; columns of $N_{H,\rm abs}=10^{22}/10^{23}$ cm$^{-2}$ have $\Delta C\geq+2133/+503$ respectively; line-forming radii of $R_\xi/R_{\rm ph}=1.1/1.3$ are disfavored by $\Delta C=+967/+496$; and, along the degenerate $\epsilon_X f_\Omega$ axis, products of $0.01/0.1/0.5$ give $\Delta C=+324/+278/+503$. As the full model evaluation is prohibitively expensive, we do not attempt to directly fit for these parameters. Our coarse grid only provides rough constraints on their characteristic values.
The blackbody radius implied by the peak luminosity and temperature, $R_{\rm bb}\approx2\times10^{11}$ cm, is much smaller than the inferred $R_{\rm ph}\sim10^{14-15}$ cm, and reconciling the two requires the emergent spectrum to be substantially harder than a blackbody, $T_{\rm col}/T_{\rm eff}=(R_{\rm ph}/R_{\rm bb})^{1/2}\sim20$--$70$. This aligns with previous expectations of photon starvation in the QPE ejecta; for instance, in the radiation-transport simulations of \citet{Vurm25}, free-free emission and Comptonization behind the shock cannot produce enough photons for a Planckian spectrum, so the radiation emerges with fewer photons of higher average energy and a temperature well above the blackbody value at the same luminosity. Electron scattering may further harden the observed spectrum. The cumulative effect is that $R_{\rm bb}$ underestimates the true emitting radius by a large factor, as discussed in \citet{Vurm25}, and $R_{\rm ph}$ rather than $R_{\rm bb}$ is the relevant scale of the emitting region.

Given the degeneracy between $f_\Omega$ and $\epsilon_X$, our data are only sensitive to the \textit{product} $f_\Omega\,\epsilon_X\sim 0.05$. An independent (though also degenerate) constraint arises from the emission line flux scaling as $\propto f_{\Omega}\,N_{H,\rm em}$ (Fig.~\ref{fig:model_evol}), where $N_{H,\rm em}/N_{H,\rm abs}$ can be a factor of several, but likely not much larger. Thus while we have anchored our estimate at $f_\Omega\sim0.5$, in principle the range $f_\Omega\sim0.25-1$ is plausible, with corresponding ranges $\epsilon_X\sim0.05-0.2$ and $N_{H,\rm em}/N_{H,\rm abs}\sim 3-12$. In the following, we propagate these uncertainties to the derived physical constraints in Section~\ref{sec:discussion} by normalizing to the fiducial value $f_{\Omega,0.5}=f_\Omega/0.5$.

\subsection{Rapid variability} \label{subsec:variability}

\begin{figure*}
    \centering
    \includegraphics[width=\textwidth]{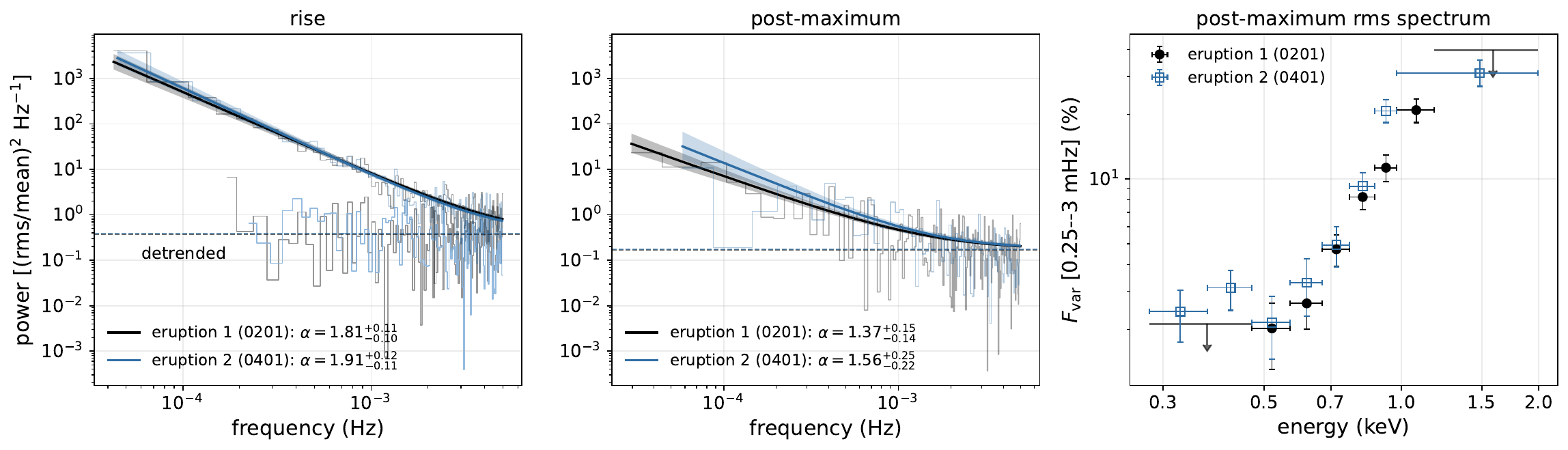}
    \caption{Rapid variability diagnostics of the two eruptions (OBSID 0201/0401 are in black/blue). \textbf{Left/center:} PSDs of the rise/peak segments fit with power-law models. The dashed lines indicate the Poisson noise floor. For the rise light curves, we also show the PSD of the detrended light curve, indicating no variability is present apart from the power-law rise. \textbf{Right:} Energy-resolved fractional variability in the 0.25--3 mHz band for the peak segments. The variability increases at higher energies.}
    \label{fig:psd}
    \label{fig:fvar_energy}
\end{figure*}

Another interesting feature of the data is the presence of stochastic variability in the luminosity following the eruption peak (Fig.~\ref{fig:lc} bottom right), which is typically not seen given the timescales and observational cadence of most QPEs (though see \citealt{Baldini26} for another example of intra-flare variability). To quantify this, we used Fourier timing methods on the light curves separated by rise and peak (Section~\ref{sec:methods}). In Fig.~\ref{fig:psd}, we show the power spectral densities (PSDs) of the rise and post-peak light curves with power-law fits, as well as the energy-resolved rms spectra after the peak. The rise PSDs are fully described by a smooth power-law increase. If we detrend the rise light curve by subtracting the best-fit power-law and recompute the PSD, all variability disappears, leaving the PSD consistent with pure Poisson noise at all frequencies above 0.25 mHz. However, after the peak, stochastic flickering is detected above the noise floor at $\Delta\ln\mathcal{L}=417/157$ for the two eruptions, with $\alpha=1.37^{+0.15}_{-0.14}/1.56^{+0.25}_{-0.22}$ and fractional variabilities of $F_{\rm var}=2.9\pm0.2\%$/$3.3^{+0.4}_{-0.3}\%$ in the 0.25--3 mHz band. The rms variability rises smoothly as a function of energy, from $\lesssim$3\% in the soft band to $\approx$10--20\% at 0.9--1.0 keV. Variability is present down to timescales of $\sim 1$ ks, which is comparable to the diffusion time at the photosphere (bottom panel of Fig.~\ref{fig:wind_profiles}), indicating the variations may arise from stochastic fluctuations in photons escaping from its outer edge.

It is also worth noting that this phenomenology is similar to the X-ray variability of thermal TDEs, which also show stochastic flickering on similar minute-long timescales with power-law PSDs (though their slopes are shallower), fractional variability at the $\lesssim 10\%$ level, and rms spectra rising steeply with energy \citep{Chakraborty26b}. In TDEs, this behavior is thought to originate from rapid temperature fluctuations in a compact, newly formed turbulent accretion flow, whose observed amplitudes are exponentially enhanced in the Wien tail \citep{Mummery25b}. Its emergence during the QPE peak and decline suggests a similar mechanism may drive the variability, involving percent-level rapid fluctuations in the photosphere temperature after the peak. They may also be driven by local fluctuations in $N_H$ or $\log\xi$ not captured in the smoothly-evolving model of Section~\ref{subsec:physical_model}; distinguishing between the causes would require hydrodynamical simulations of the wind launching mechanism.

\section{Discussion} \label{sec:discussion}

Fig.~\ref{fig:schematic} presents a schematic of the geometry and physical scales implied by our model of a time-evolving wind kinetically powering the eruption luminosity while simultaneously producing the evolving spectral absorption lines (Section~\ref{sec:results}). The eruptions are powered by wide-angle, mildly relativistic mass loss in a wind-like outflow with a velocity $v_w\approx0.2c$. The outflow launching radius, $R_{\rm launch}$, can be estimated by setting the escape speed there to equal $v_w$:
\begin{equation}
    R_{\rm launch} = 50\,R_g\;v_{w,0.2c}^{-2} \approx 7\times10^{12}\;\mathrm{cm}\;M_{\bullet,6}\,v_{w,0.2c}^{-2}
    \label{eq:rlaunch}
\end{equation}
Our model implies the photosphere radius begins from this value, then steadily increases to $R_{\rm ph}\sim10^{14-15}$ cm as the continuum temperature climbs to $k_BT_{\rm ph}\gtrsim 100$ eV at eruption peak. The dominant absorption occurs close to the photosphere, at the line-forming radius $R_\xi\sim1.2R_{\rm ph}$, with a column density $N_H\sim5\times10^{22}$ cm$^{-2}$. The emission component of the P Cygni profile is redshifted by a magnitude comparable to the absorption blueshift, suggesting a biconical or asymmetric outflow geometry.
 
In contrast to the homologous expansion interpretation of \cite{Chakraborty25b}, our new higher-quality data and analytical model reveal that the evolution in absorption/emission can be modeled by the ionization response within a time-evolving wind without invoking significant changes in the bulk outflow velocity. In reality, both effects are likely to contribute, but this would require more sophisticated time-dependent radiative transfer modeling to ascertain.

\subsection{Constraints on energetics, long-term evolution, and QPE models} \label{subsec:energetics_models}

Our model estimates a wind kinetic power of $\dot E_K\sim 10L_X\,f_{\Omega,0.5}$, which tracks the luminosity by assumption of a constant radiative efficiency. Each eruption ejects an integrated $M_{\rm ej}\sim 10^{-3}\,f_{\Omega,0.5}\,M_\odot$, radiating $E_{\rm rad}\sim4\times10^{48}$ erg in total with a corresponding kinetic energy $E_K\sim 4\times10^{49}\,f_{\Omega,0.5}$ erg. 

While our model assumes the QPEs are kinetically powered via the wind-like outflow rather than direct accretion onto the SMBH, as the radiated luminosity scales as $\propto\dot{M}_{\rm out}v_w^2$ rather than $\dot{M}_{\rm in}c^2$, we can still constrain models in which the QPEs are powered by episodes of transient enhanced accretion onto the SMBH \citep[e.g.][]{Pan23,Dodd25} or a secondary orbiting black hole \citep[e.g.][]{Lam25,Pelle26}. In such models, the mass required to power a single burst is:
\begin{equation}
    M_{\rm acc}= \frac{\eta_XE_{\rm rad}}{\eta_{\rm acc} c^2}\approx2\times10^{-4}\,M_\odot\;\eta_{\rm acc,0.1}^{-1}\;\eta_{X,10}
\end{equation} 
where $\eta_{\rm acc}$ is the radiative efficiency of accretion and $\eta_X$ is the bolometric correction from X-rays. $M_{\rm acc}$ thus appears significantly smaller than the mass ejected in the outflow. For comparison, \cite{Tombesi12} find the opposite for ultrafast outflows in high-$\dot{M}_{\rm Edd}$ AGN, which have $\dot{M}_{\rm out}/\dot{M}_{\rm acc}\sim 0.05-0.1$. This strains accretion-based models to power the QPEs due to the large observed $\dot{M}_{\rm out}/\dot{M}_{\rm acc}\gtrsim 5f_{\Omega,0.5}$ (though see \citealt{Xrism25}). While there is uncertainty in $\eta_X$, it is likely not large enough to account for this discrepancy, as constraints on the QPE broadband SED suggest most radiation is emitted close to the peak temperature \citep{Wevers25}. Perhaps most importantly, the ability of the wind model to kinetically power the radiation and simultaneously reproduce the spectral line evolution disfavors accretion-powered models, as it is unclear whether they can produce the same result.

If we assume QPEs are powered by orbiter-disk collisions in an EMRI, our data provide some new constraints on the nature of the interaction. The near-identical consecutive bursts (Fig.~\ref{fig:lc}) favor models in which the QPEs are observed either once per orbit, or with two bursts so close together that they overlap to create a single observed QPE. The latter would require a significantly eccentric orbit, an idea we revisit below. The former may be due to asymmetry in the mass of the forward and backward ejecta from the collisions \citep{Huang25,Jankovic26}; inability of the diffuse, low-density debris streams to penetrate the denser accretion disk \citep{Linial25,Yao26b}; an eccentric EMRI with an apocenter larger than the disk outer edge; and/or occultation of one eruption parity by a warped disk geometry \citep{Mummery26a}. In this model, the outflows should be launched close to the shock velocity of the collision, which is comparable to the Keplerian velocity of the orbiter at pericenter:
\begin{equation}
    v_p \approx 0.03c\;M_{\bullet,6}^{1/3}\,P_{11\,\rm d}^{-1/3}\,\bigg(\frac{1+e}{1-e}\bigg)^{1/2}
\end{equation}
For $e=0$, this is a factor of several smaller than the $v_w\sim0.2c$ required, though this may partially arise from a large velocity spread in the ejecta \citep[e.g.][]{Huang25,Jankovic26,Yao26b}. Matching $v_p=v_w$ requires $e\approx0.95$, which, suggestively, is \textit{also} roughly the eccentricity required by setting $r_p=R_{\rm launch}$ (Eq.~\ref{eq:rlaunch}). Such a highly eccentric and tight orbit is technically possible, as it narrowly avoids tidal disruption for a solar-mass star, but the formation channel of such an orbit warrants investigation. Moreover, with $a\approx1000\,R_g$ and $e\approx0.95$, the implied apsidal precession period is $\approx32.5\,P_{\rm orb}$. Given that our \textit{XMM} observations are of two consecutive QPEs, the lack of such effects is not surprising, but there may be systematic modulations in the outflow and/or burst properties over the full \textit{NICER} campaign, which we defer to future work.

The mass ejections may also perturb the orbit itself. If the ejecta are launched with a net velocity component $u_{\rm ej}$ anti-parallel to the orbital motion in the orbiter's rest frame, momentum conservation requires $M_*\,\Delta v_\parallel \sim f_\parallel\,M_{\rm ej}\,u_{\rm ej}$, where $f_\parallel$ is the momentum-weighted anisotropy. Since the orbital period depends only on the specific orbital energy, and an impulse changes that energy by $\mathbf{v}\cdot\Delta\mathbf{v}$, the fractional period change per orbit is:
\begin{align}
    \frac{\Delta P}{P} &\approx 0.02\;f_{\parallel,0.2}\,M_{\bullet,6}^{-1/3}\,u_{\mathrm{ej},0.2c}\,P_{\rm 11\;d}^{1/3} \nonumber \\
    &\qquad\times\left(\frac{M_{\rm ej}/M_*}{10^{-3}}\right) \left(\frac{\sqrt{(1+e)/({1-e)}}}{6.2}\right)
\end{align}
where the above estimate is anchored at $e=0.95$. This is in fact comparable to the observed $\dot{P}\approx(1.70\pm0.02)\times10^{-2}$ d d$^{-1}$ \citep{Chakraborty26a}; however, it is unclear whether the required anisotropy can be produced in the stream fed by stellar debris, motivating further study of the collision dynamics (see also \citealt{Linial24a} for further discussion).

The inferred mass-loss also sets the allowed lifetime of the eruptions. At $\sim 10^{-3} M_\odot$ per burst, a Sun-like orbiter powering the eruptions with its own mass cannot sustain significantly more than $\mathcal{O}(1000)$ bursts before being destroyed, implying an upper bound of $\lesssim 30\,f_{\Omega,0.5}^{-1}$ years for QPE activity in Ansky for a mean recurrence time of 11 days. If we instead consider two mass-loss events per orbit, even though only one is observable in X-rays, the bound halves to $\lesssim 15\,f_{\Omega,0.5}^{-1}$ years. The bound would further decrease if the star progressively loses more mass with each collision \citep[e.g.][]{Yao25}, though the magnitude of this effect is uncertain and would require hydrodynamical simulations to estimate. For a direct star-disk collision the ejected mass is approximately $M_{\rm ej}\sim\pi\,R_*^2\,\Sigma$, where $\Sigma\approx10^8\,f_{\Omega,0.5}$ g cm$^{-2}$ is the implied disk surface density. This is unphysically high, implying the eruptions are unlikely to be powered by collisions with a ``hard sphere'' and may instead require extended stream-disk collisions \citep[e.g.][]{Linial25}. These bounds apply only if the ejecta are drawn from the orbiter; if they are predominantly shocked disk gas---as for stellar-mass black hole impactors, or in extended stream--disk collisions---the reservoir is the disk and no such bound applies. The constraints derived from our observations thus provide a new tool for making concrete predictions and distinguishing between QPE models.

\subsection{Implications for reverberation and feedback} \label{subsec:feedback_reverberation}

Integrated over an eruption, the wind carries $E_K\gtrsim 10^{49}\,f_{\Omega,0.5}$ erg and $10^{-3}\,f_{\Omega,0.5}\,M_\odot$ into the circumnuclear medium, for a cumulative $\gtrsim 10^{51}\,f_{\Omega,0.5}$ erg and $\sim0.1\,f_{\Omega,0.5}\,M_\odot$ ejected at mildly relativistic speeds thus far. This material should shock against the ambient gas and radiate synchrotron emission in the radio, in direct analogy to the delayed radio afterglows observed from TDE outflows \citep[e.g.][]{Goodwin23,Cendes24,Alexander26}. Because every eruption ejects material at a similar $v_w$, the successive shells expand together as a thick, quasi-continuous blast wave. We follow the treatment of \citet{Matsumoto21} and \citet{Linial26} and normalize to fiducial values of $E_{K,51}=E_K/10^{51}$ erg, $n_{\rm cnm,1}=n_{\rm cnm}/10^1$ cm$^{-3}$, $\epsilon_{e,-1}=\epsilon_e/10^{-1}$, and $\epsilon_{B,-2}=\epsilon_B/10^{-2}$ for the fractions of shock energy in non-thermal electrons and magnetic fields, and adopt an electron index $p=2.5$. All kinetic energy and mass budgets scale linearly with $f_\Omega$ (Section~\ref{subsec:physical_model}); we quote them for $f_\Omega=0.5$. The ejecta free-expand until they sweep up a mass comparable to their own, decelerating after:
\begin{equation}
    t_{\rm dec} \simeq \bigg(\frac{3M_{\rm ej}}{4\pi n_{\rm cnm}m_pv_w^3}\bigg)^{1/3} \approx 4.6\,\mathrm{yr}\;E_{K,51}^{1/3}n_{\rm cnm,1}^{-1/3}v_{w,0.2c}^{-5/3}
\end{equation}
at which point the radius reached by the wind will be $R_{\rm dec}\approx v_wt_{\rm dec}\approx 10^{18}\,\mathrm{cm}\;E_{K,51}^{1/3}n_{\rm cnm,1}^{-1/3}v_{w,0.2c}^{-2/3}$. The peak flux density at this time is roughly:
\begin{equation}
    F_{\nu}^{\rm max} \approx 18\,\mathrm{mJy}\;\epsilon_{e,-1}\epsilon_{B,-2}^{0.88}n_{\rm cnm,1}^{0.88}E_{K,51}v_{w,0.2c}^{1.75}\nu_{\rm obs,3\,GHz}^{-0.75}d_{104}^{-2}
\end{equation}
where $d_{104} = d/104\;\rm Mpc$, the luminosity distance normalized to Ansky's distance. This implies a delayed radio peak a few years after QPE onset; the uncertainty is dominated by $n_{\rm cnm}$ and the shock microphysics ($\epsilon_e/\epsilon_B$). For comparison, equipartition modelling of the late-time radio flares has inferred outflow energies reaching $\sim10^{51}$ erg in a large fraction of TDEs \citep{Cendes24,Alexander26}, so Ansky's cumulative budget already sits at the top of that range at a distance typical of the detected sample.

Observationally, Ansky is undetected thus far in the radio, with a deepest upper limit of $<0.45$ mJy at 3 GHz in July 2024 \citep{Hernandez25a}, while the systematic radio survey of twelve QPE hosts conducted in \cite{Goodwin25} found that seven (including Ansky) showed no detectable counterpart. However, those data do not yet test the prediction above, given that X-ray limits place the flare onset around or before early 2024. The results and modeling we present imply that continued radio monitoring may eventually detect an afterglow signature if the nuclear density is sufficient.

Dust reprocessing into infrared wavelengths offers a more prompt, if less energetic, channel. For instance, infrared echoes of TDEs are observed to radiate up to $\sim10^{51}$ erg, comparable to the budget available here \citep{Masterson24}. Balancing grain heating by the eruption peaks against thermal re-emission, the sublimation radius for graphite grains is roughly:
\begin{equation}
    R_{\rm sub}\approx 0.06\,\mathrm{pc}\;L_{43.5}^{1/2}\,a_{0.1}^{-1}\,T_{1850}^{-2.0}
\end{equation}
where $a_{0.1}$ is the grain size in units of $0.1\,\mu$m and $T_{1850}$ is the dust sublimation temperature in units of $1850$ K. The light-crossing time of this shell, $R_{\rm sub}/c\approx 12$ d, is comparable to the recurrence period, so the dust cannot resolve individual eruptions. The reprocessed luminosity is $L_{\rm IR}\sim f_{\rm dust}\langle L\rangle$, with covering fractions ranging from $f_{\rm dust}\sim0.01$ in optically selected TDEs to $\gtrsim0.1$ in dusty nuclei \citep{Masterson24}, giving $\nu F_\nu\approx10^{-14}-10^{-12}$ erg cm$^{-2}$ s$^{-1}$ near the $2-5\,\mu$m peak for $1500$ K grains, well within reach of current facilities. However, the signal may be suppressed: JWST spectroscopy of the host shows optically thin silicate emission inconsistent with a standard optically thick torus \citep{Sanchez26}, and the slow mid-infrared rise seen by WISE lagging the 2019 optical transient by $\sim$one year \citep{Sanchez24} suggests the pre-existing dust was already excavated, so little may survive near $R_{\rm sub}$. The ejecta themselves reach $R_{\rm sub}$ within a few months at $v_w\sim0.2c$ and may destroy the innermost grains. The former would delay any echo to the light-crossing time of the surviving dust, while the latter would produce a slowly declining rather than brightening mid-infrared excess; in either case, late-time infrared photometry is well-motivated to test these predictions.

It is finally worth placing these numbers alongside the outflows usually invoked for SMBH feedback. Time-averaging the integrated kinetic energy per burst yields a ``steady-state'' $\langle\dot E_K\rangle\approx 4\times10^{43}$ erg s$^{-1}$. For $M_\bullet\sim10^6\,M_\odot$, this is $30\%$ of the Eddington luminosity, significantly larger than the $\dot E_K/L_{\rm bol}\sim0.5$--$5\%$ measured for ultrafast outflows in high-$L_{\rm Edd}$ AGN, the regime generally regarded as sufficient for efficient coupling to the host \citep{DiMatteo05,Tombesi12}. In absolute terms, the cumulative $\gtrsim 10^{51}$ erg exceeds the gravitational binding energy of the gas within the central parsec ($\sim10^{47}$ erg for $n=10$ cm$^{-3}$) by four orders of magnitude, so the cumulative kinetic energy is energetically sufficient to strongly perturb or clear gas within the inner parsec, depending on the coupling efficiency. This would in turn lower the ambient density $n_{\rm cnm}$ entering the deceleration timescale above, and lengthen the expected delay to any radio flare. Any such clearing must be confined to the inner parsec, since the ionized gas across the host remains kinematically cold and carries no imprint of past outflows. The comparison can also be drawn against the nucleus itself: the ionizing luminosity it sustained over the preceding $\gtrsim1500$ yr integrates to a similar $\sim10^{51}$ erg \citep{Sanchez26}, so a couple of years of eruptive activity have supplied as much energy in kinetic form as the accretion flow radiated over millennia. Compared with TDEs, whose outflows liberate $10^{48-51}$ erg in a one-off event, Ansky delivers a comparable budget within roughly a year of eruptive activity and then persists. QPEs with energetic budgets comparable to Ansky thus constitute a quasi-continuous channel for depositing energy and mass in galactic nuclei.

In summary, while no infrared or radio reverberation signatures associated directly with the QPEs have been observed yet, we encourage continued monitoring as our results indicate that the eruption energies are more than sufficient to produce bright multiwavelength signatures (up to $\sim 18$ mJy in radio and $\sim10^{-12}$ erg cm$^{-2}$ s$^{-1}$ in infrared) delayed by a few years if the circumnuclear medium is dense enough.

\section{Conclusion} \label{sec:conclusion}

We presented and analyzed new \textit{XMM-Newton} observations of the rise-to-peak in two consecutive QPEs of ZTF19acnskyy/``Ansky'' (Fig.~\ref{fig:lc}), which exhibits the brightest, longest-duration, most energetic QPEs to date \citep{Hernandez25a}. Our data provide the deepest view into the individual eruptions of any QPE, which we use to make new constraints on the physical processes generating the bursts. With time-resolved X-ray spectroscopy, we detect a time-varying P Cygni profile arising from L-shell resonance transitions of highly ionized iron (Fe \textsc{xix}--\textsc{xxiv}) beginning in the earliest rise phases and evolving in centroid, depth, and width throughout the burst evolution in response to the changing continuum. This dramatic evolution on such short timescales is unprecedented in SMBH outflows.

To interpret the data, we constructed a time-dependent analytical toy model (Section~\ref{subsec:physical_model}) of a wind-like outflow of constant velocity ($v_w$) turning on from $\dot{M}_w=0$ to kinetically power the luminosity under the assumption of a constant radiative efficiency. In this model, the density profile in the wind is set by the mass outflow rate at the retarded time, $\rho(r,t)\propto \dot{M}(t-r/v_w)$. This choice is necessitated by the rapid evolution of the light curve and spectra in Ansky, in contrast to the usual steady-state continuity assumption used in AGN/XRB wind modeling. The model provides a self-consistent description of the ionization response and spectral evolution of the wind powering the eruptions (Figs.~\ref{fig:model_evol}--\ref{fig:wind_profiles}).

While numerically evaluating the full model is too expensive for parameter fitting, we evaluate a coarse grid of models to estimate the wind properties as $v_w\sim0.2c$, X-ray production efficiency $L_X/\dot{E}_K\sim 0.05-0.2$, and covering fraction $f_\Omega\sim 0.25-1$, implying outflow rates of $\dot{M}\sim 10^{-9}-10^{-8}\,M_\odot$ s$^{-1}$ for a per-eruption integrated mass budget of $\sim10^{-3}\,f_{\Omega,0.5}\,M_\odot$, kinetic energy budget $\gtrsim10^{49}\,f_{\Omega,0.5}$ erg, and radiated energy $\gtrsim10^{48}$ erg. The good agreement between the model and both the light curve and spectral data provides strong evidence for the QPE emission arising from a kinetically powered wind-like outflow rather than direct accretion onto the SMBH. This picture agrees qualitatively with recent work modeling QPEs arising from the collisions of extended debris streams surrounding an EMRI orbiter with the SMBH accretion disk \citep{Linial25,Yao26b}; however, it is not yet clear whether other models (e.g. disk instabilities) are entirely \textit{unable} to produce such behavior, making definitive model comparisons premature.

At these energetics, the material already ejected should shock against the circumnuclear matter and may produce delayed reverberation at radio and/or infrared wavelengths if the gas/dust content of the nucleus is sufficiently high, motivating future multiwavelength observing campaigns to find these signatures. Moreover, at $\langle\dot E_K\rangle/L_{\rm bol}\sim 0.3$, the QPEs in Ansky represent a contribution to black hole feedback comparable to that from AGN outflows \citep[e.g.][]{Tombesi12}, and may play a significant role in shaping the nuclear environment of the host galaxy, though this depends on the uncertain QPE formation rate and lifetime \citep[e.g.][]{Arcodia24b,Chakraborty25a}. Our preliminary estimates of kinetic feedback in Ansky motivate further study of whether QPEs may provide a significant feedback mechanism over a typical galaxy lifetime.

Our measurements also provide a new tool for testing physical models of QPEs. For instance, if we assume QPEs are powered by star-disk collisions in an EMRI, the measured mass-loss rate implies the eruptions can persist for at most $\sim10^3$ further eruptions if the ejecta are drawn from a stellar orbiter, and that the orbit must have significant eccentricity ($e\sim0.95$) to produce the large wind velocities near pericenter passage. Furthermore, the implied mass-loss rate and ejecta velocities may even account for the large positive $\dot{P}\sim10^{-2}$ d d$^{-1}$ \citep{Chakraborty26a} if ejected anisotropically, resulting in velocity kicks to the orbiter near each pericenter passage. The near-identical light curves and spectra between the two consecutive bursts also disfavor models which produce two well-separated observable bursts per orbit, suggesting instead reproducible collision dynamics and only one QPE per orbital period. Our modeling estimates the physical scales, densities, and kinematics of ejecta launched by individual eruptions---quantities that hydrodynamic simulations of orbiter-disk collisions now predict directly \citep[e.g.][]{Huang25,Jankovic26,Yao26b}---establishing a new observational probe of QPE models orthogonal to recent tests involving QPE timings \citep{Miniutti25,Chakraborty26a,Arcodia26}. The simple analytical model presented here is deliberately empirical, and provides a remarkably good description of the data for relatively few free parameters. Its success motivates both a more refined treatment of the radiation-hydrodynamic processes in the wind launching and subsequent evolution, as well as similar analyses across the QPE population to compare their underlying physical mechanisms.

\section*{Acknowledgments}

We thank the \textit{NICER} and \textit{Swift} teams for high-cadence monitoring of ZTF19acnskyy. We thank the \textit{XMM-Newton} team for coordinating with us to schedule and then re-schedule these observations, even as Ansky's burst timings defied all reasonable expectations in a cosmic attempt to thwart our campaign \citep{Chakraborty26a}. We thank Philippe Yao, Andy Mummery, Paul Draghis, and Megan Masterson for useful discussions which contributed to this work, and the anonymous referee for an insightful report which improved the manuscript.

IL acknowledges support for this work provided by NASA through the NASA Hubble Fellowship grant \#HST-HF2-51581.001-A awarded by the Space Telescope Science Institute, which is operated by the Association of Universities for Research in Astronomy, Inc., for NASA, under contract NAS5-26555. PK acknowledges support by NASA through the NASA Hubble Fellowship grant HST-HF2-51534.001-A awarded by the Space Telescope Science Institute. BDM acknowledges support from NASA (grant 80NSSC24K0934) and NSF grant (AST-2406637).  The Flatiron Institute is supported by the Simons Foundation. LHG acknowledges financial support from ANID program FONDECYT Iniciaci\'on 11241477. EQ acknowledges support through the European Space Agency (ESA) Research Fellowship in Space Science.  MG is funded by Spanish MICIU/AEI/10.13039/501100011033 and ERDF/EU grant PID2023-147338NB-C21.

\bibliography{refs}{}
\bibliographystyle{aasjournal}

\appendix
\counterwithin{figure}{section}
\counterwithin{table}{section}

\section{Supplemental figures} \label{app:supplemental_figures}

\begin{figure}
    \centering
    \includegraphics[width=\linewidth]{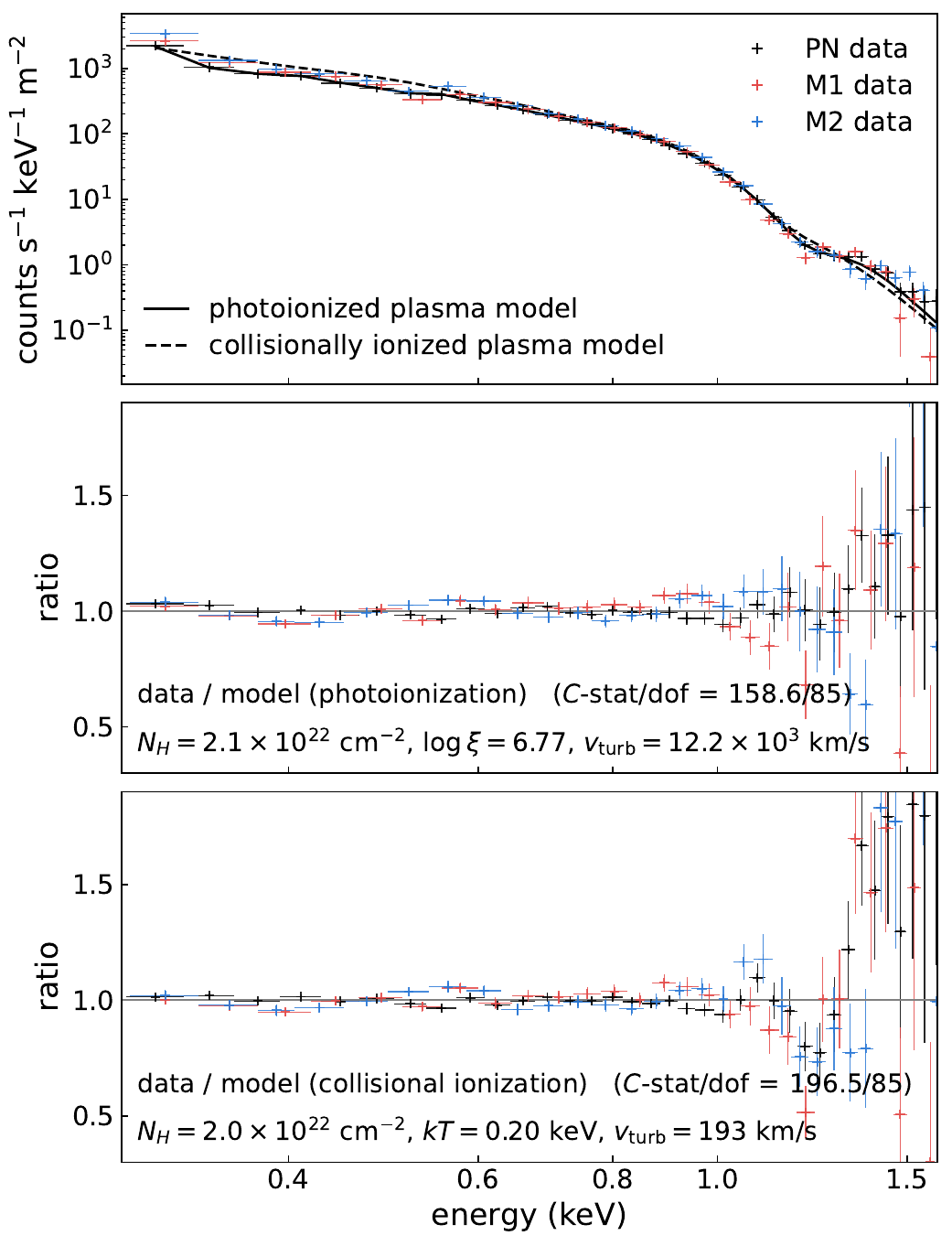}
    \caption{Photoionized versus collisionally ionized absorber fits to a
    representative high-SNR window near the peak of the first eruption:
    the EPIC spectra with both best-fit models overlaid (top) and the
    data-to-model ratios for each (middle, bottom), with best-fit
    parameters and statistics annotated.}
    \label{fig:hot_vs_pion}
\end{figure}

In Fig.~\ref{fig:hot_vs_pion}, we compare a representative high-SNR spectrum to models of absorption by optically thick photoionized and collisionally ionized plasmas (both with a Gaussian component for the emission). The collisional ionization model fails to reproduce the depth and width of the $\sim$1.0--1.2 keV Fe-L trough, resulting in a significantly worse fit statistic ($\Delta C=+37.9$). The same holds for most of our time-resolved spectra (Fig.~\ref{fig:model_pref}).

\begin{figure}
    \centering
    \includegraphics[width=\linewidth]{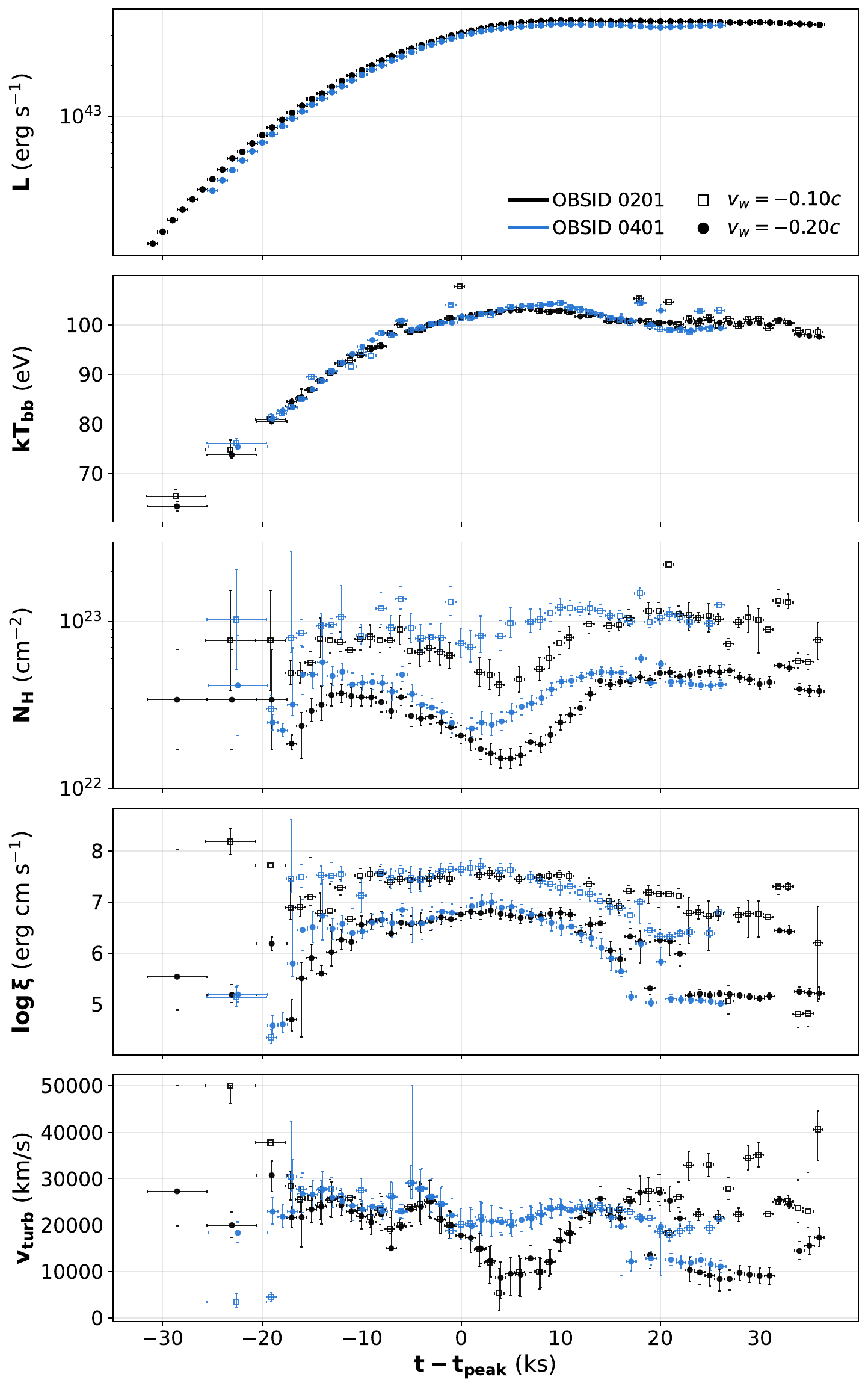}
    \caption{Photoionized plasma absorption fits to the time-resolved spectra of both eruptions, with the absorbing column, ionization parameter, and turbulent velocity free in every window and the bulk outflow velocity frozen to $-0.1c$ (open squares) or $-0.2c$ (filled circles). All error bars are $\Delta C=1$. The earliest low-count phases use the combined windows, whose $N_H$ was frozen to the per-eruption median column and is plotted with a $\pm0.3$ dex systematic error bar.}
    \label{fig:time_indept_evol}
\end{figure}

In Fig.~\ref{fig:time_indept_evol} we show the results of fitting each time-resolved spectrum \textit{independently}, i.e. with no relationship of the absorber $N_H$ or $\log\xi$ between windows. We show both models with frozen $v_w=\{0.1c,\,0.2c\}$ to visualize the degeneracy effect on $\log\xi$. For the earliest low-SNR spectra, we froze $N_H$ to the median of each eruption and plot a $0.3$ dex systematic error, as the data at those epochs are not sensitive enough to constrain the parameter. The data broadly prefer $N_H\sim10^{22-23}$ cm$^{-2}$, $\log\xi\sim 5-7.5$ erg cm s$^{-1}$, and $v_{\rm turb}\lesssim 0.1c$, consistent with the parameters preferred by the full time-dependent wind model of Section~\ref{subsec:physical_model}. The formal \texttt{SPEX} $1\sigma$ errors depicted in these plots likely significantly underestimate the true uncertainty, particularly in $\log\xi$; however, rather than choose an arbitrary systematic uncertainty, we show the error bars as is.

\end{document}